\documentclass[useAMS,usenatbib]{mn2e}
\usepackage{longtable}
\usepackage{threeparttable}
\usepackage{subfigure}
\usepackage{graphicx}
\usepackage{float}

\title[A lucky imaging multiplicity study of exoplanet host stars]{A lucky imaging multiplicity study of exoplanet host stars}
\author[C. Ginski et al.]{C. Ginski$^{1}$\thanks{E-mail:ginski@astro.uni-jena.de}, M. Mugrauer$^{1}$, M. Seeliger$^{1}$, and T. Eisenbeiss$^{1}$\\
$^{1}$ Astrophysikalisches Institut und Universit\"ats-Sternwarte Jena, Jena 07743, Germany}

\begin{document}

\pagerange{\pageref{firstpage}--\pageref{lastpage}} \pubyear{2011}

\maketitle

\label{firstpage}

\begin{abstract}
To understand the influence of additional wide stellar companions on planet formation, it is necessary to determine the fraction of multiple stellar systems amongst the known extrasolar planet population.    \\
We target recently discovered radial velocity exoplanetary systems 
observable from the northern hemisphere and with sufficiently high proper motion to detect stellar companions via direct imaging. 
We utilize the Calar Alto 2.2\,m telescope in combination with its lucky imaging camera AstraLux.\\ 
$71$ planet host stars have been observed so far, yielding one new low-mass ($0.239 \, \pm \, 0.022 \, M_{\odot}$) stellar companion, $4.5 \,$arcsec (227\,AU of projected separation) northeast of the planet host star HD\,185269, detected via astrometry with AstraLux. We also present follow-up astrometry on three previously discovered stellar companions, showing for the first time common proper motion of the $0.5 \,$arcsec companion to HD\,126614.
Additionally, we determined the achieved detection limits for all targets, which allows us to characterize the detection space of possible further companions of these stars.    \end{abstract}


\section{Introduction}

Recent radial velocity (RV) and transiting exoplanet surveys have resulted in the discovery of a large number of new exoplanetary systems.
Most stars are born in binary or multiple systems (\citealt{b10}, \citealt{b11}), and indeed we observe that a reasonable fraction of the stars in the Galaxy are multiple (\citealt{b1}). The effects of additional stellar companions on the planet formation process are therefore of high interest to constrain and calibrate planet formation theories.\\
A number of studies have been conducted on this subject in the past, such as described in \cite{b12} or \cite{b14} and most recently in \cite{b15} and \cite{b16}. As a result of these studies, $43$ multiple stellar systems hosting exoplanets are known to date, suggesting that about $17$ \% of all known exoplanets reside in such systems (\citealt{b13}).\\
In this paper we present the results of our ongoing multiplicity study of the most recently (since 2008) discovered RV exoplanet host stars. We utilized the Calar Alto 2.2\,m telescope and the AstraLux instrument (\citealt{b5}) to achieve imaging data with higher Strehl ratio compared to simple seeing-limited imaging.\\ 
In the following section we characterize our sample. 
In section \ref{sec-obs} we give a brief introduction to the observation technique as well as the instrument used and describe the reduction and astrometric calibration of our data. In section \ref{results} we show all confirmed or rejected companion candidates with the associated proper motion analysis. We also present detection limits for all studied systems. Finally in section \ref{conclusions} we summarize our findings.

\section{Target sample}

\begin{figure}
\includegraphics[scale=0.355]{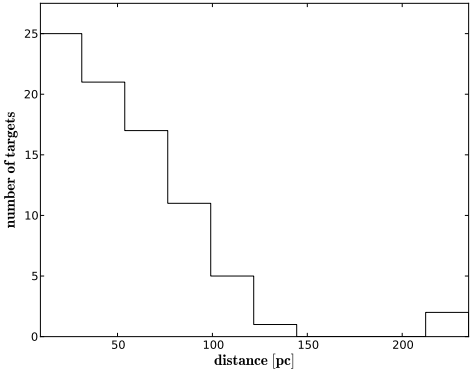}
\caption[]{Histogram of distances of all our target stars (as listed in exoplanet.eu by \citealt{b29}). The majority being within 100\,pc.}
\label{fig:dist-hist}
\end{figure}

\begin{figure}
\includegraphics[scale=0.4]{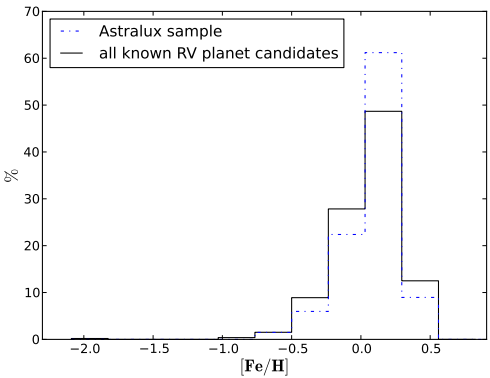}
\caption[]{Metallicity distribution for all targets in our sample. For comparison we plot the metallicity distribution of all RV planet candidate host stars discovered so far (as listed in exoplanet.eu by \citealt{b29}).}
\label{fig:metall}
\end{figure}

\begin{figure}
\includegraphics[scale=0.355]{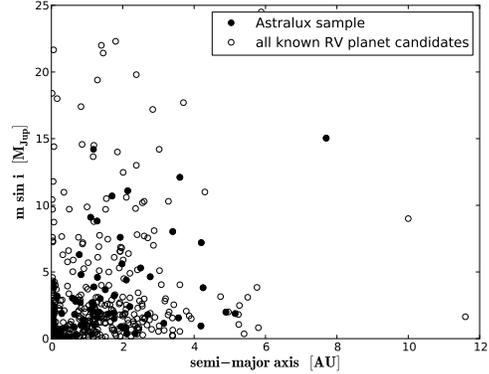}
\caption[]{Minimum mass versus separation for all planet candidates orbiting hosts included in our sample (as listed in exoplanet.eu by \citealt{b29}). For comparison we also plot minimum mass versus separation of all RV planet candidates discovered so far.}
\label{fig:mass-sep-plot}
\end{figure}

\begin{figure}
\includegraphics[scale=0.4]{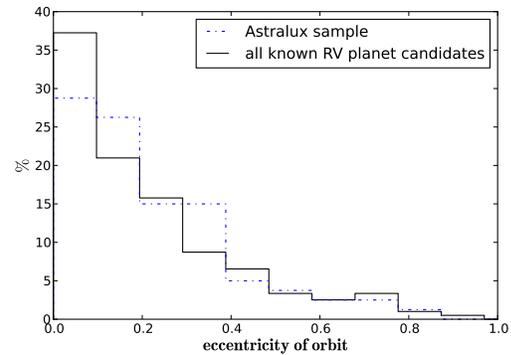}
\caption[]{Eccentricity distribution of planet candidates orbiting hosts included in our sample. For comparison we also plot the eccentricity distribution of all RV planet candidates discovered so far (as listed in exoplanet.eu by \citealt{b29}).}
\label{fig:e}
\end{figure}

Our sample consists of stars with RV planet candidates discovered between 2008 and 2011. They are all observable from the northern hemisphere with declinations down to -22$^{\circ}$ and a relatively even distribution in right ascension. Since we want to test companionship of detected companion candidates via astrometry, we chose targets which are close enough to show a sufficient proper motion (e.g. $>$ 47\,mas/yr or approximately one Astralux pixel per year) to do a common proper motion analysis approximately one year after the first epoch images were taken. The distance distribution of our targets is shown in Fig.~\ref{fig:dist-hist}. On average our targets are 56.6\,pc away and show a proper motion of 180.4\,mas/yr equivalent to 3.8 pix/yr on the Astralux instrument.\\
The RV planet search technique favors main-sequence stars, therefore the average age of our target stars is 4.6\,Gyr (data as listed in exoplanet.eu by \citealt{b29}). Spectral types range from late F to early M, with the majority being G and K type. In Fig.~\ref{fig:metall} we show the metallicity distribution of our sample compared to the whole RV planet candidate host population. The average metallicity of our sample is slightly higher than solar metallicity at 0.09. The metallicity distribution of our sample does not differ significantly from the whole population.\\ 
In Fig.~\ref{fig:mass-sep-plot} we present the minimum masses of detected RV planet candidates versus their semi-major axis for our sample and the whole population. The majority of the detected planet candidates are located within 2\,AU of their hosts. The average minimum mass of our sample is 2.8\,M$_{Jup}$, as compared to 2.4\,M$_{Jup}$ for the whole RV planet candidate population. Given the standard deviation of 3.3\,M$_{Jup}$ and 3.8\,M$_{Jup}$ respectively this is not significant.\\
The eccentricity distribution of RV planets around our sample stars can be seen in Fig.~\ref{fig:e}. We compared the eccentricity distribution of our sample to that of all RV planet candidates and found no significant differences.\\  
We conclude that the general properties of our sample resemble those of the whole population. 

\section{Observations outline and data calibration}
\label{sec-obs}
\subsection{Lucky imaging technique}

To achieve high spatial resolution as well as high sensitivity, one must consider the effects of turbulent atmosphere. Given the Rayleigh criterion for the minimum resolvable angle of $ \Theta = 1.22 \cdot \lambda \cdot D^{-1}$, the Calar Alto 2.2m telescope has a theoretical resolution of $89\,$mas at a wavelength of $776\,$nm, which is the central wavelength of the SDSS \textit{i}' filter which we are using. The median seeing at the Calar Alto site, however, is $0.9\,$arcsec (\citealt{b4}), which is about $10$ times larger.\\
The atmospheric conditions in the optical regime are subject to rapid variations with a timescale that can be approximated by the speckle coherence time $\tau_e \approx 0.36 \cdot r_0 \cdot \Delta\nu^{-1}$ (\citealt{b7}), where $r_0$ is the Fried-parameter and $\Delta\nu$ is the wind speed dispersion in the atmosphere. For a typical wind speed dispersion of $\Delta\nu \approx 10 \,$m$\cdot$s$^{-1}$ at the Calar Alto site and a V-band seeing of $0.7 \,$arcsec, $\tau_e$ is in the order of $100\,$ms (\citealt{b8}).
Within $\tau_e$ the resulting speckle pattern will remain fixed, whereas longer integration times would lead to an averaged and therefore "smeared out" speckle pattern.\\
The lucky imaging approach consists of taking several thousand short images with integration times shorter than $\tau_e$, to sample the speckle variations during the observation window. We then only choose the so called "lucky shots" with a very high Strehl ratio in one of the speckles, to shift and add, resulting in a final image with the highest possible strehl ratio and therefore highest possible angular resolution.\\  
For an in-depth introduction to the lucky imaging technique please see \cite{b6}. 
 
\subsection{Observations and data reduction}

All observations were carried out with the Calar Alto 2.2\,m telescope in combination with the AstraLux instrument. 
This instrument consists of a back-illuminated, electron-multiplying, frame transfer CCD, which is well suited for the lucky imaging observation technique. For a detailed description of the instrument, see \cite{b5}.\\
Since we wanted to detect low-mass stellar or even brown dwarf companions of our targets, 
we chose the SDSS \textit{i}' filter as described in \cite{b25} for all observations. This is the best choice taking into account the sensitivity of the detector and the brightness of the detectable companions, as well as the variability of the atmosphere. For calibration we took dome- and sky-flats at the beginning and/or end of each observation night.\\
For all science targets we chose $29.54\,$ms of exposure time per frame, which is well below the typical speckle coherence time at the Calar Alto site. Shorter exposure times would have led to significantly growing overheads since we would have needed to switch off the frame transfer mode of the instrument. Longer exposure times would have resulted in a less localized and more "smeared out" speckle pattern.\\
The electron-multiplying gain was individually adjusted for each target to obtain the maximum signal while still operating in the linear regime of the detector. We then took 100 dark frames with closed shutter and with the respective electron multiplying settings before the start of each acquisition sequence.\\
Depending on weather conditions and time constraints we took between 10000 and 80000 frames per science target, which corresponds to a total integration time of $0.25\,$min and $1.97\,$min respectively for a typical frame selection rate of $5\,$\%.   
The obtained AstraLux images are all reduced and processed with our own ESO-MIDAS (Munich Image Data Analysis System) software for the reduction of imaging data taken with the lucky imaging technique. The individual AstraLux images are dark-subtracted with the median (for optimal suppression of cosmic ray artifacts) of several short integrated dark-frames and flat-fielded. For the lucky imaging data processing, the reached Strehl-Ratio in all individual images is determined at first using the bright planet host star as Strehl-probe. The frames are then ranked according to their Strehl-Ratio and only the best $10$, $5$, and $1\,$\% of all frames, i.e. only the frames with the highest Strehl-Ratio, are then selected and combined using the shift+add technique. Thereby, for image registration, the position of the brightest pixel in the speckle pattern of the planet host star is determined in all frames, which are then shifted and averaged.

\subsection{Astrometric calibration}

Since we want to precisely measure separation and position angle of our companion candidates with respect to the primary stars, and the AstraLux instrument is not permanently mounted to the Calar Alto 2.2m telescope, a careful astrometric calibration of the detector in each observation epoch is necessary. For this purpose we selected a sample of five wide binaries, for which precise Hipparcos measurements as well as several additional observation epochs are available. We show our sample along with the Hipparcos astrometry in Table~\ref{tab: binaries}. Furthermore, we observed the center of the globular cluster M15 whenever possible, for which precise HST astrometry is available.\\
For each epoch we have a minimum of three calibrators which were observed in the same night as the science targets. For all calibrators we used the Hipparcos measurement of epoch $1991.25$ as reference point and then linearly fitted the slow orbital motion by using all data points available in the Washington Double Star (WDS) catalog.\\
The pixel scales and position angles of the calibration images for each individual calibrator were calculated using ESO-MIDAS for the position measurements of the binary components, and GAIA (Graphical Astronomy and Image Analysis Tool) for the respective measurements of the M15 cluster components (using SExtractor: Software for source extraction by \citealt{b26}).  
The average of these calculations were used as final calibration for the respective epoch to cancel out 
systematic errors due to residual orbital motion of our binary stars. The final astrometric calibrations for all observation epochs are listed in Table~\ref{tab: astrocal}.

\begin{table}
  \caption{Hipparcos astrometry of epoch 1991.25 of our calibration binaries}
  \begin{tabular}{@{}ccc@{}}
  \hline

 Binary 		& Separation $[arcsec]$	&  Position Angle $[^\circ]$\\
 \hline
 HIP\,59585		&	$18.677 \pm 0.032$		&	$190.83 \pm 0.10$					\\
 HIP\,65205		&	$15.041 \pm 0.015$		&	$219.44 \pm 0.06$					\\
 HIP\,67099		&	$18.751 \pm 0.028$		&	$343.33 \pm 0.09$					\\
 HIP\,72508		&	$15.248 \pm 0.014$		&	$91.36 \pm 0.05$					\\
 HIP\,80953		&	$16.356 \pm 0.021$		&	$195.69 \pm 0.07$					\\
 \hline\end{tabular}

\label{tab: binaries}
\end{table}

\begin{table}
 \caption{Astrometric calibration of all observation epochs. We list the pixel-scale (PS) and the position angle (PA) of the y-axis for all observation epochs.}
 \begin{threeparttable}
  \begin{tabular}{@{}ccc@{}}
  \hline 
 Epoch 	& PS $[mas/pix]$&  PA of y-axis $[^\circ]$\\
 \hline
 23/04/2008		&	$47.278\pm 0.101$	&	$0.17 	\pm 0.23$		\\
 11/07/2008		& $47.180\pm 0.078$	& $0.44 	\pm 0.15$		\\
 16/01/2009		& $47.237\pm 0.114$	& $359.39 \pm 0.08$		\\
 07/09/2009		& $47.220\pm 0.100$	& $0.43		\pm 0.13$		\\
 23/02/2010		& -\tnote{1}	&	-\tnote{1}		\\
 14/07/2010		& $47.243\pm 0.049$	& $358.49 \pm 0.22$		\\
 14/01/2011		& $47.365\pm 0.135$	& $358.07 \pm 0.12$		\\
 27/07/2011		& $47.160\pm 0.066$	& $358.37	\pm 0.23$		\\
\hline\end{tabular}
\label{tab: astrocal}
\begin{tablenotes}\footnotesize 
\item[1] due to bad weather conditions no astrometric calibrators could be observed
\end{tablenotes}
\end{threeparttable}
\end{table}

\subsection{PSF subtraction and astrometric measurements}

Since we are searching for faint low-mass stellar companions to bright stars, we subtract the primary stars' PSF  in order to detect the companions with highest possible signal-to-noise. Due to the random nature of the lucky imaging approach, the PSFs in the final reduced images may vary significantly from star to star, for example as a function of the average zenith distance of the observed objects. Hence it is not feasible to subtract a PSF standard from all the images. Furthermore, the PSFs of our target stars are not necessarily symmetrical in nature, rendering simple rotation subtraction techniques ineffective.\\
Given these limitations, we found the best approach to be an unsharp mask filter. We fold the images with a 2-dimensional Gaussian of size $n$ to "blur" them, and then subtract the "blurred" images from the originals, thus eliminating all low spatial frequencies from the images. We tested sizes $n$ between $5$ and $20$ pixels for different images, with $n\,=\,10$ generally yielding the best results in terms of improving the signal-to-noise of the faint companion candidates.\\
We again used ESO-MIDAS for all astrometric measurements, first measuring the primary star's position in the original images, then applying the PSF subtraction as described, before measuring the companion candidate's position. We always measured both positions multiple times, averaging the results to avoid statistical errors.     

\section{Results}
\label{results}

To date we observed $71$ planet host-stars in our project. In seven cases we detected faint companion candidates, of which three were already known. In the following paragraphs, we show the astrometric measurements for all the companion candidates, as well as the proper motion analysis, to determine which of the objects are physically associated to the target stars. 

\subsection{Imaging of known companions}

\begin{figure*}
\subfigure[$\tau$ Boo taken in epoch 23/04/2008]{
\includegraphics[scale=0.295]{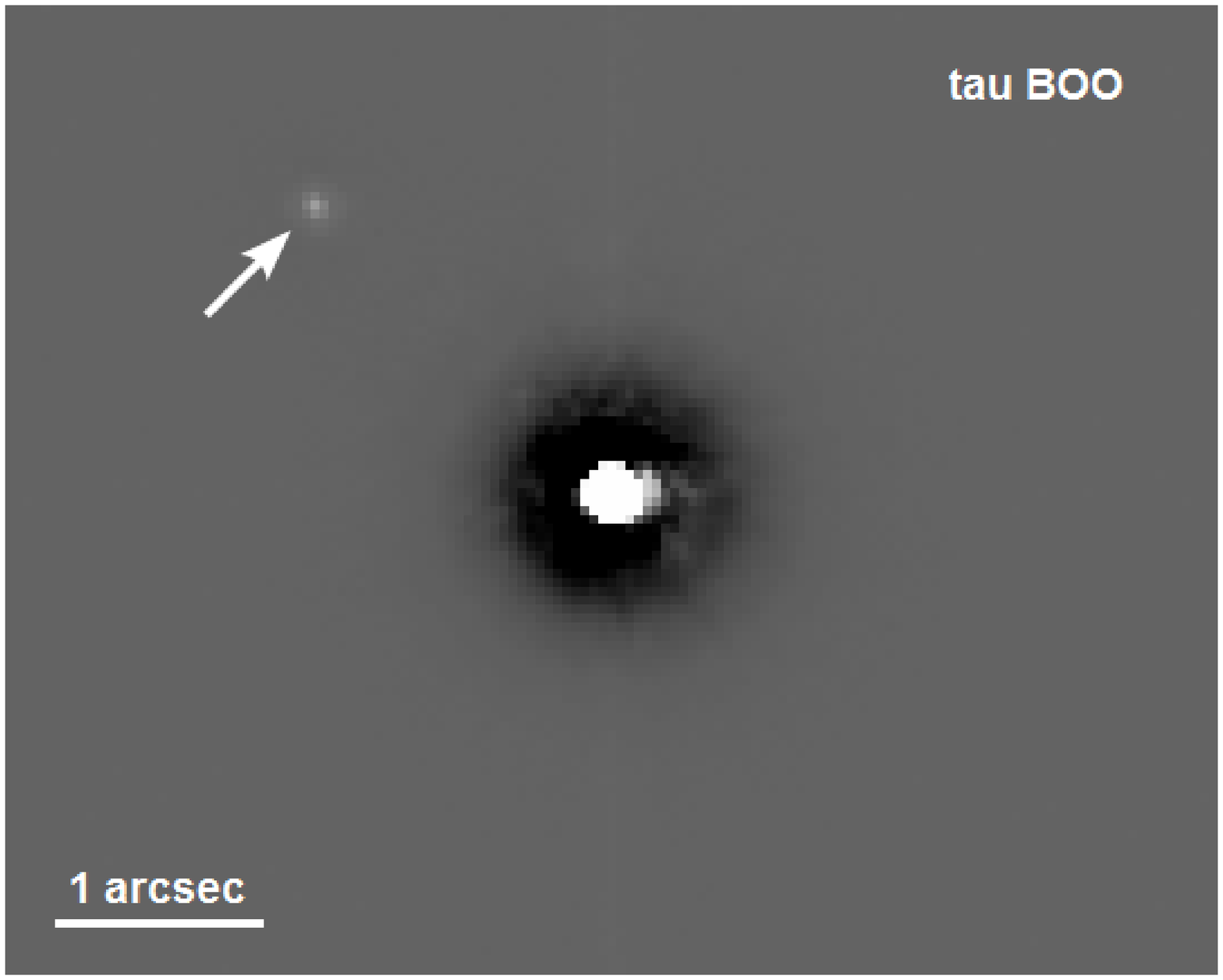}
\label{tauboo}
}
\subfigure[HD\,176051 taken in epoch 27/07/2011]{
\includegraphics[scale=0.295]{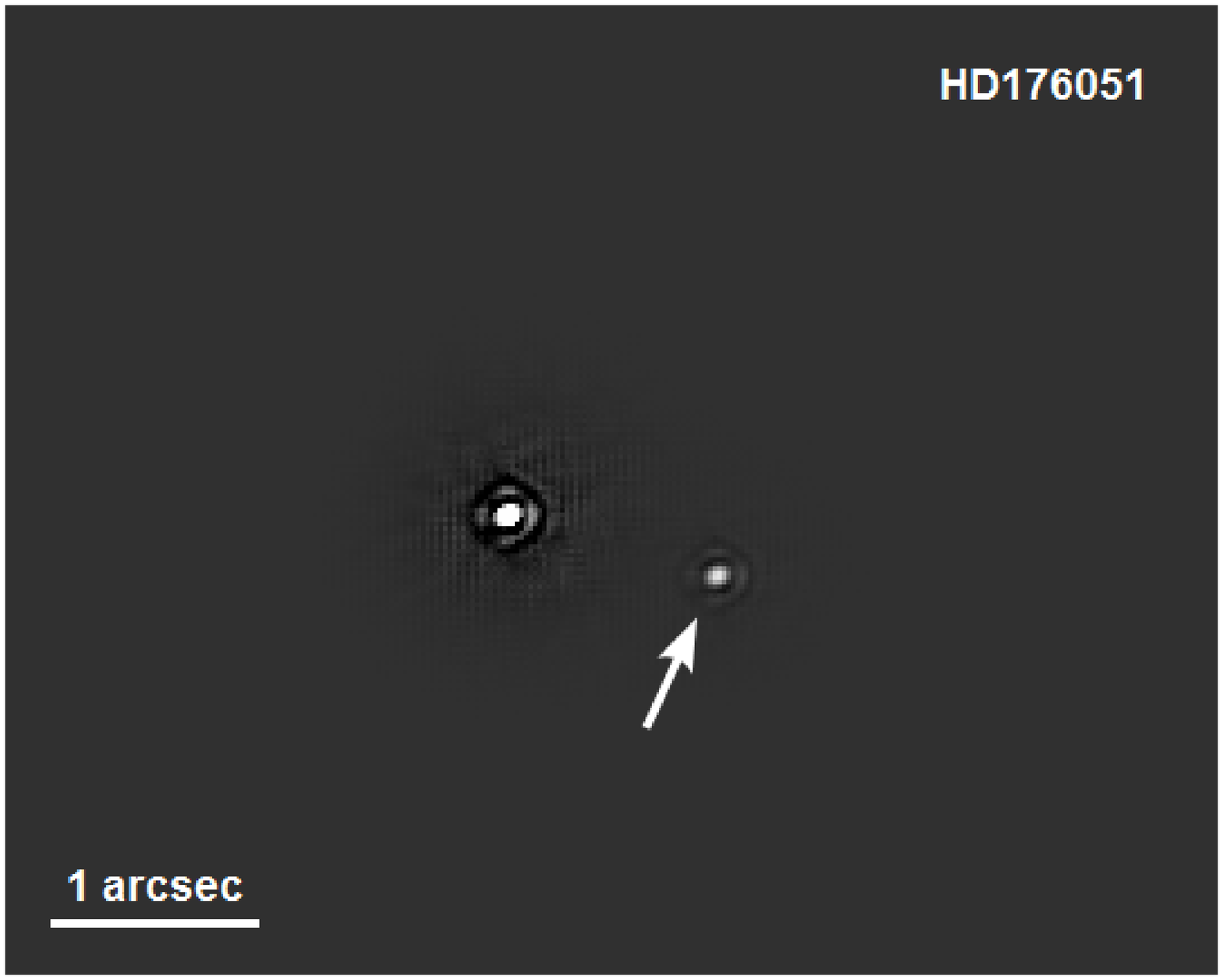}
\label{hd176051}
}
\subfigure[HD\,126614 taken in epoch 14/01/2011]{
\includegraphics[scale=0.295]{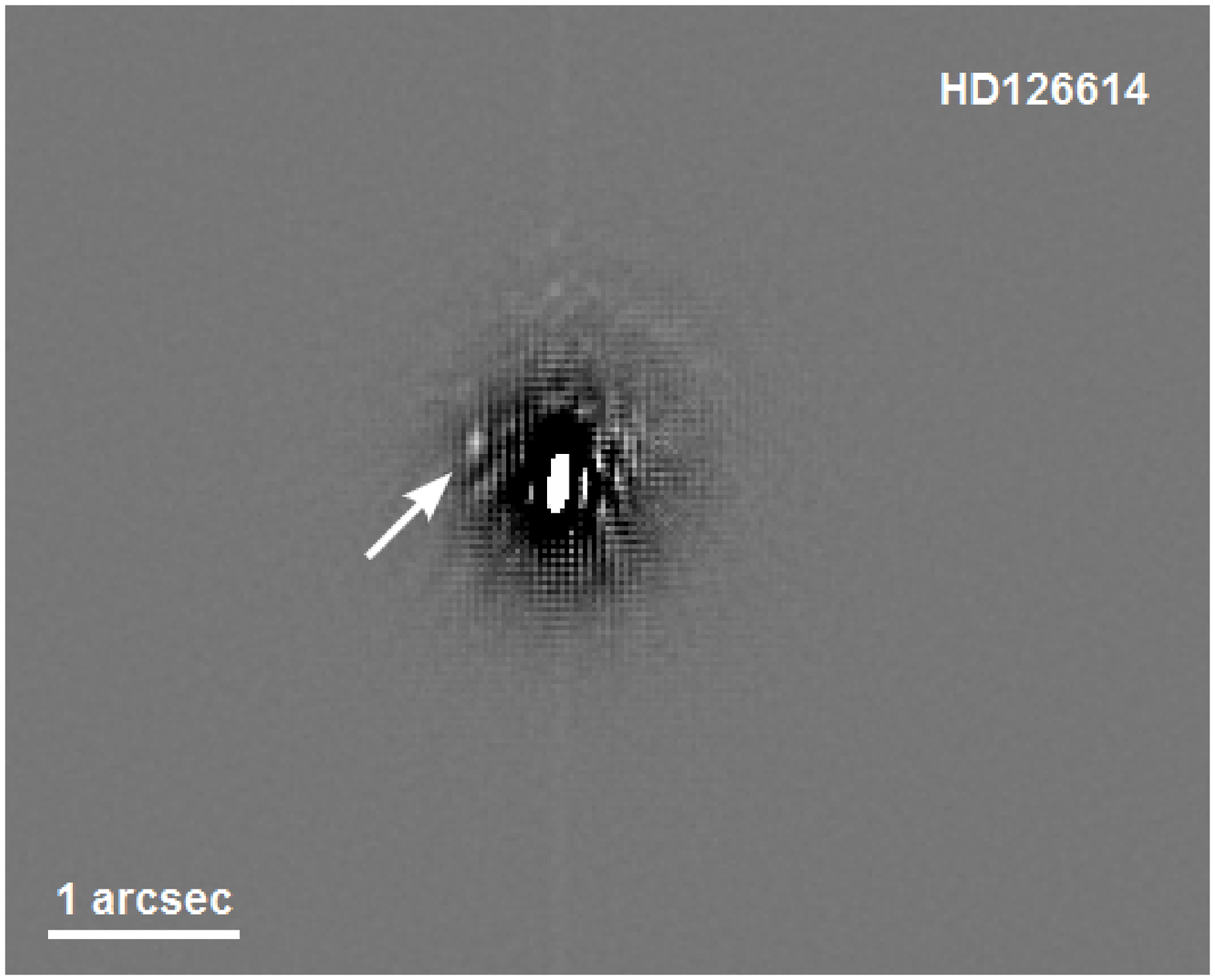}
\label{hd126614}
}

\caption[]{Follow-up images of already known companions to exoplanet host stars taken in SDSS \textit{i}' filter during our ongoing campaign. All images have been PSF subtracted. North is always up and east to the left. The companions are marked by arrows.} 
\label{fig:known}
\end{figure*}

In the course of our study we imaged three already known companions to exoplanet hosts, to test the instrument capabilities and to provide additional astrometric measurements for future orbit solutions. The final reduced images can be seen in Fig.~\ref{fig:known} and our astrometric measurements are listed in Table~\ref{tab: known}. \\
The stellar companion to $\tau$ Boo resides at an angular separation $2.181 	\pm	0.018\,$arcsec, which corresponds to a projected separation of $34\,$AU. \cite{b19} stated, that $\tau$ Boo\,A and B form an eccentric binary system with a semi-major axis of $\approx 225\,$AU, and masses of $\approx 1.3 \, M_{\odot}$ and $\approx 0.4 \, M_{\odot}$ respectively.\\
HD\,176051\,A and B form a close binary system of only $1.139	\pm 0.009 \,$arcsec separation and an orbital period of $61.2\,$yr. The stellar masses are $1.07 \, M_{\odot}$ and $0.71 \, M_{\odot}$ respectively (\citealt{b18}). \cite{b18} found an astrometric companion of $\approx 1.5\, M_J$, to be orbiting one of the components.\\
The stellar companion to HD\,126614 was only detected recently by \cite{b17} at the Palomar Observatory, using adaptive optics imaging. It is separated from the primary only by $0.490 	\pm 0.051\,$arcsec, which corresponds to a projected separation of $35.6 \,$AU. The mass of the companion was inferred from the JHK photometry, to be $0.324 \pm 0.004 \, M_{\odot}$. Its companionship was so far only confirmed by photometry and radial velocity trend. We are not aware of any further observations of this companion, especially regarding whether or not it is actually co-moving with the primary. Hence we present a proper motion analysis in section \ref{sec:pm}, using the measurements by \cite{b17} and our own AstraLux data point, to confirm that HD\,126614\,A and B form indeed a common proper motion pair.

\begin{table}
  \caption{Astrometric measurements of all known companions}
  \begin{tabular}{@{}cccc@{}}
  \hline
 Primary 			& Epoch 			&	Separation						&  Position Angle 			\\
 							&							&	[arcsec]							&	 $[^\circ]$						\\
 \hline
 $\tau$ Boo		&	23/04/2008	&	$2.181 	\pm	0.011$		&	$46.26 \pm	0.42$			\\
 HD\,176051		& 27/07/2011	&	$1.139	\pm 0.005$		& $252.12\pm 	0.40$ 		\\
 HD\,126614		& 14/01/2011	& $0.499 	\pm 0.067$		& $60.70 \pm 	5.60$			\\
\hline\end{tabular}
\label{tab: known}
\end{table}

\subsection{New companion candidates}

\begin{figure*}
\subfigure[Image of HD\,13931 taken in epoch 14/01/2011]{
\includegraphics[scale=0.7]{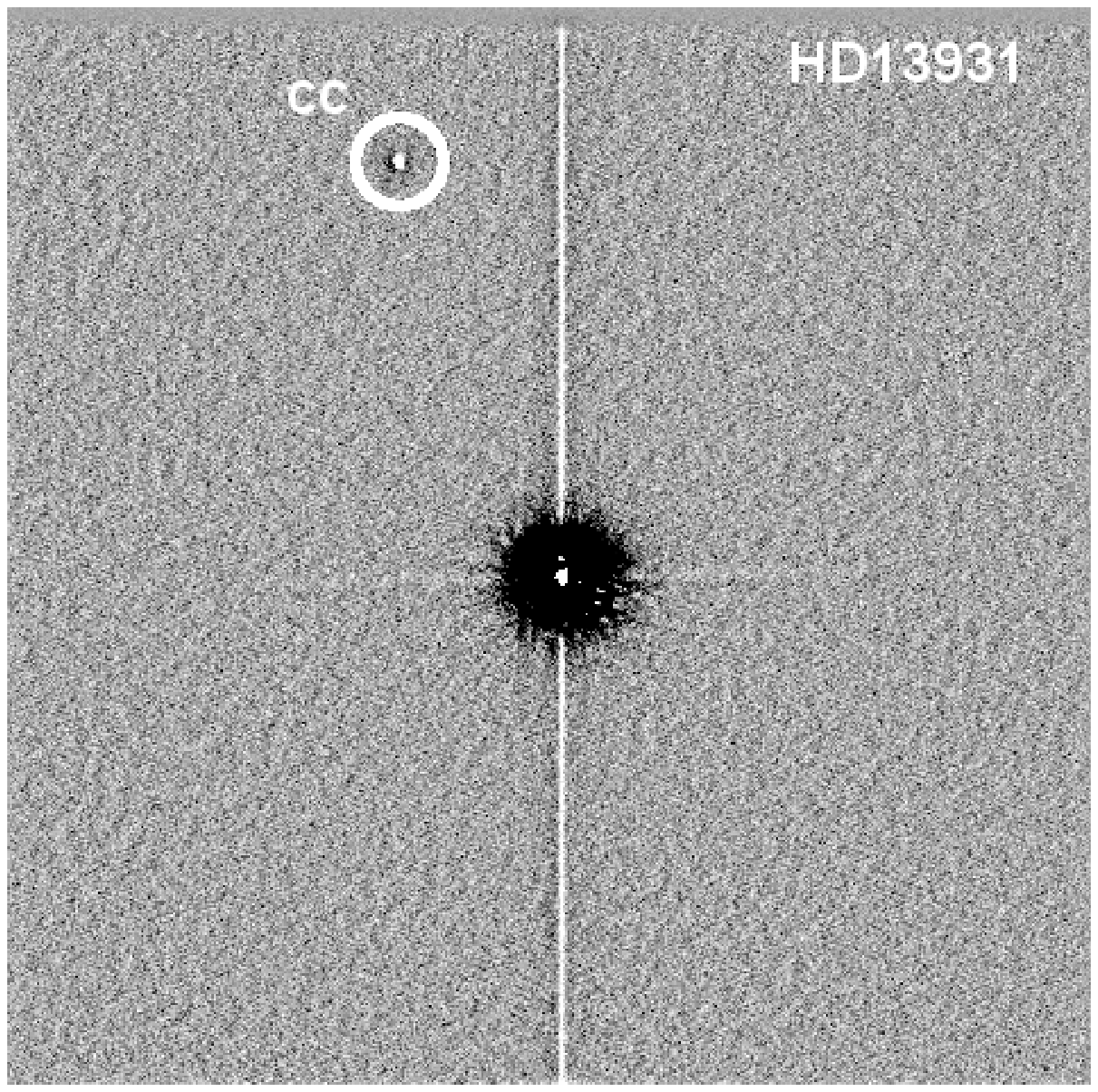}
}
\subfigure[Image of HD\,185269 taken in epoch 07/09/2009]{
\includegraphics[scale=0.7]{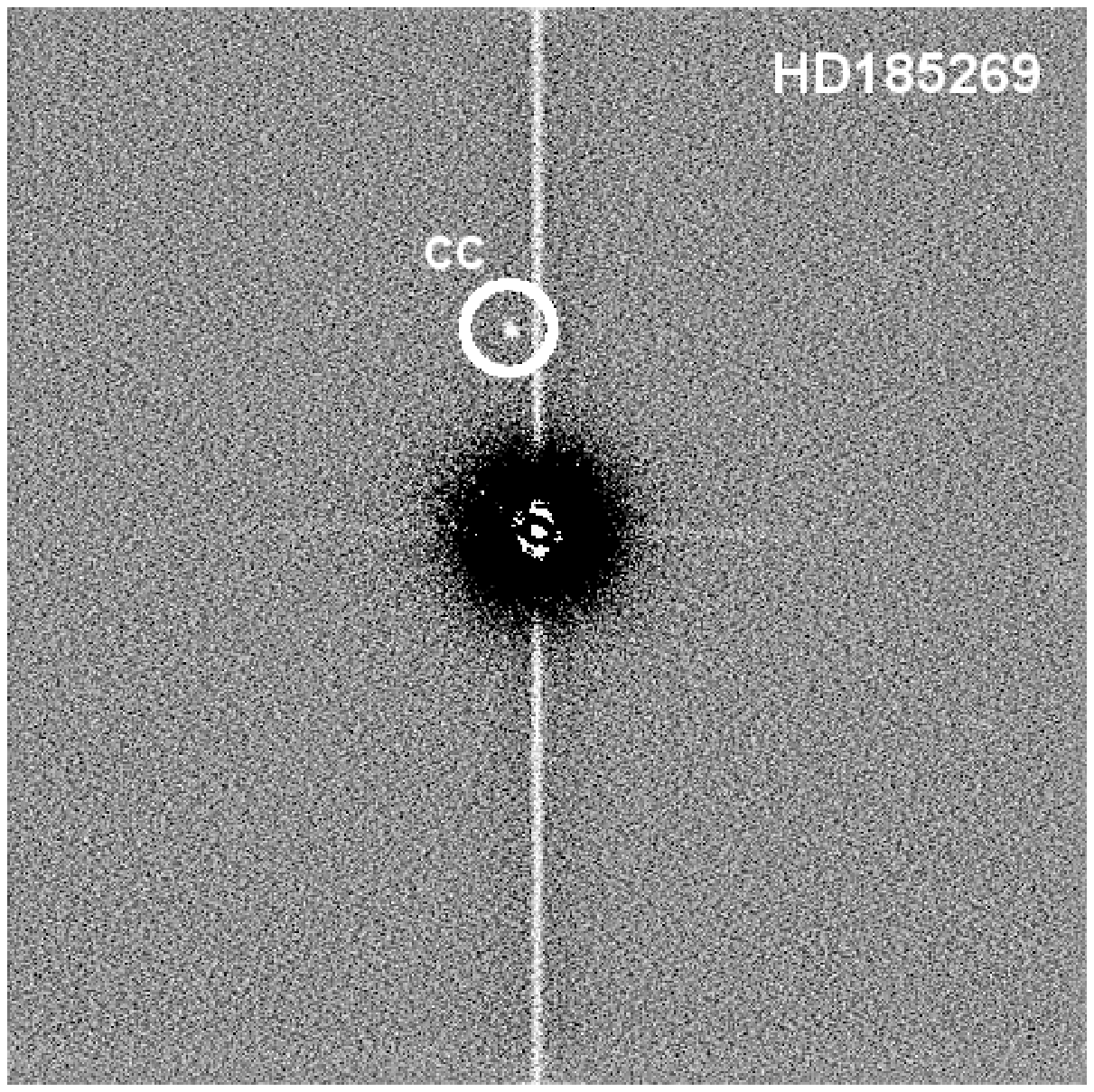}
}
\subfigure[Image of HD\,183263 taken in epoch 23/04/2008]{
\includegraphics[scale=0.7]{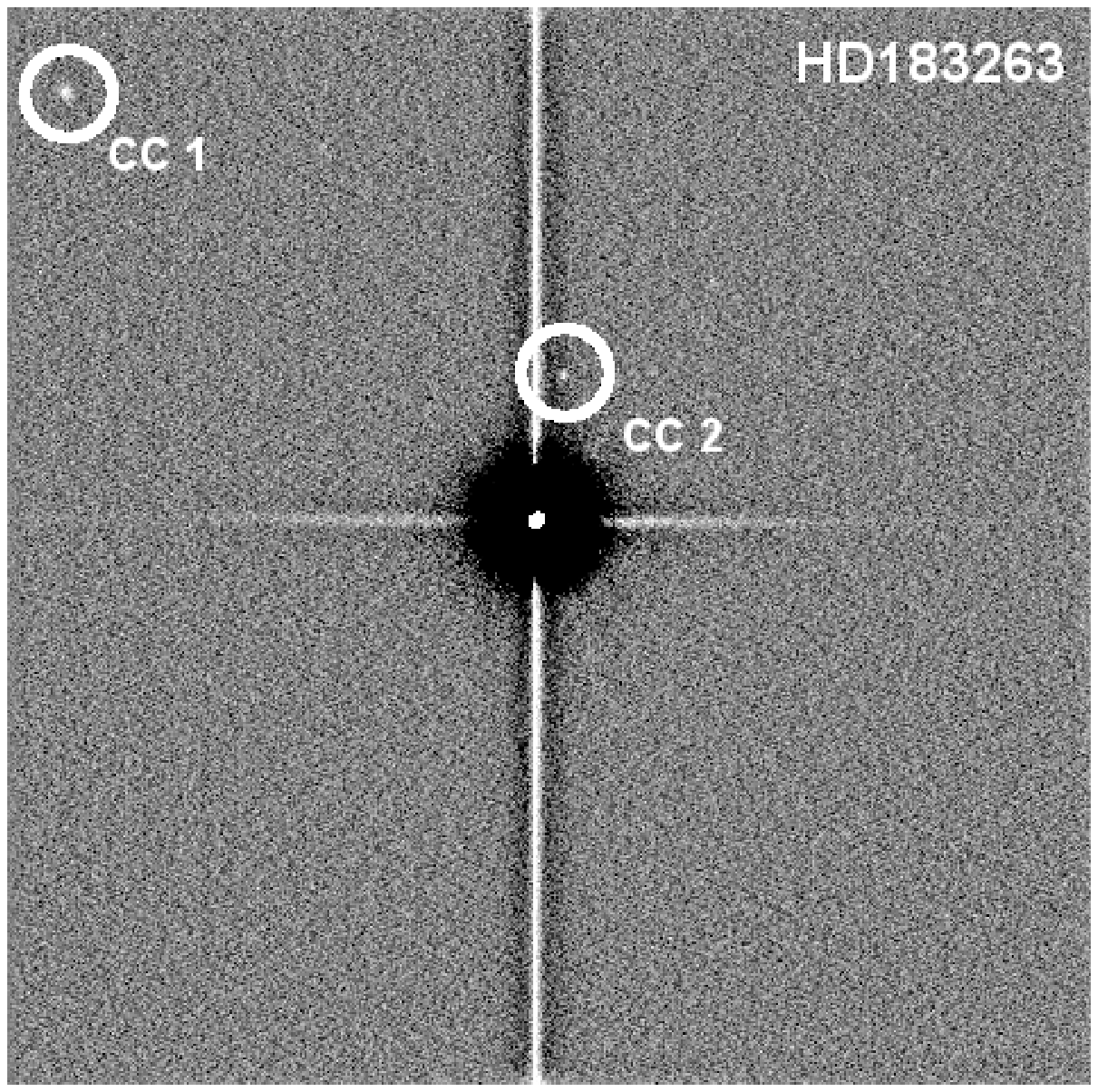}
}
\subfigure[Image of HD\,187123 taken in epoch 07/09/2009]{
\includegraphics[scale=0.7]{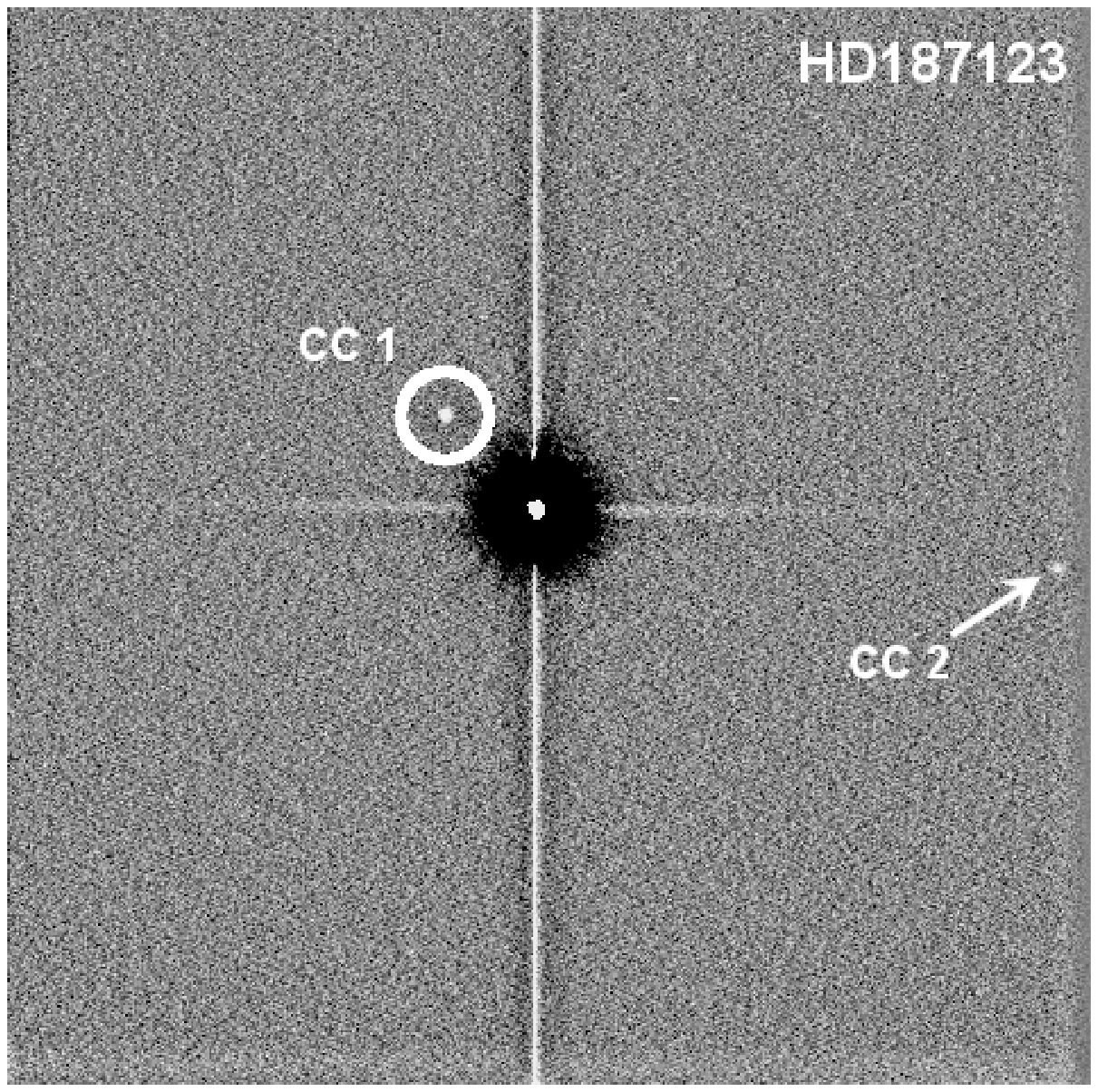}
}
\caption[]{Images of our newly discovered companion candidates taken in SDSS \textit{i}' filter. Images are $24.2\, \times \, 24.2 \,$arcsec, north is always up and east to the left. In all cases the primary stars PSF was subtracted by an unsharp mask filter. Companion candidates are marked with circles and arrows. The numbers assigned to multiple companion candidates match those used in Table~\ref{tab:cc-measurements}. }
\label{fig:cc-images}
\end{figure*}

In the course of our study we found new faint companion candidates around the planet host stars HD\,185269, HD\,183263, HD\,187123 and HD\,13931. In Fig.~\ref{fig:cc-images} we show the epochs with the highest signal-to-noise.

In Table~\ref{tab:cc-measurements} we show all the astrometric measurements obtained in the different observation epochs for all companion candidates. Given the distances of the primary stars, the detected objects would have projected separations ranging from $179\,$ AU (close candidate of HD\,187123) up to $781\,$AU (far candidate of HD\,187123). 

\begin{table*}
  
\caption{Astrometric measurements of all companion candidates}

  \begin{threeparttable}
  \begin{tabular}{@{}ccccc@{}}
  \hline

 Primary 			& Epoch 	& \# CC & Separation [arcsec]	& Position Angle $[^\circ]$\\
 \hline
  HD\,13931		&	14/01/2011	&	1			&		$9.972 \pm 0.030$			&	$19.50 \pm 0.16$ 	\\
 							&	27/07/2011	&	1			&		$10.003\pm 0.015$			&	$19.16 \pm 0.25$ 	\\
 HD\,185269		& 07/09/2009	&	1			&   $4.522 \pm 0.014$ 		& $7.91 \pm 0.23$		\\
 							& 14/07/2010	&	1			&	  $4.520 \pm 0.010$			&	$8.17 \pm 0.30$ 	\\
 							&	27/07/2011	&	1			&		$4.511	\pm 0.008$		&	$8.44	\pm 0.28$		\\
 HD\,183263		&	23/04/2008	&	1			&		$14.190 \pm 0.032$		&	$47.83 \pm 0.26$ 	\\
 							&							&	2			&		$3.295 \pm 0.015$			&	$349.55 \pm	0.40$ \\
							&	11/07/2008	&	1			&		$14.183 \pm 0.028$		&	$47.92 \pm 0.19$ 	\\
							&							&	2			&		$3.245 \pm 0.017$\tnote{1}			&	$349.72	\pm 0.31$\tnote{1} \\
 							&	07/09/2009	&	1			&		$14.244 \pm 0.033$		&	$47.93 \pm 0.17$ 	\\
 							&							&	2			&		$3.315 \pm 0.031$			&	$350.75 \pm 0.56$ \\
 HD\,187123		&	11/07/2008	&	1			&		$2.926 \pm 0.011$			&	$48.11 \pm 0.29$ 	\\
							&							&	2			&		$11.560	\pm 0.022$		&	$263.12	\pm 0.19$ \\
 							&	07/09/2009	&	1			&		$2.917 \pm 0.013$			&	$43.92 \pm 0.29$ 	\\
 							&							&	2			&		$11.696 \pm	0.028$		&	$263.92	\pm 0.17$ \\
 						
\hline\end{tabular}
\label{tab:cc-measurements}
\begin{tablenotes}\footnotesize 
\item[1] due to the spurious detection of the companion candidate this data point seems dubious
\end{tablenotes}
\end{threeparttable}
\end{table*}

\subsection{Proper motion analysis}
\label{sec:pm}

\begin{figure*}
\subfigure[Companion candidate to  HD\,13931]{
\includegraphics[scale=0.4]{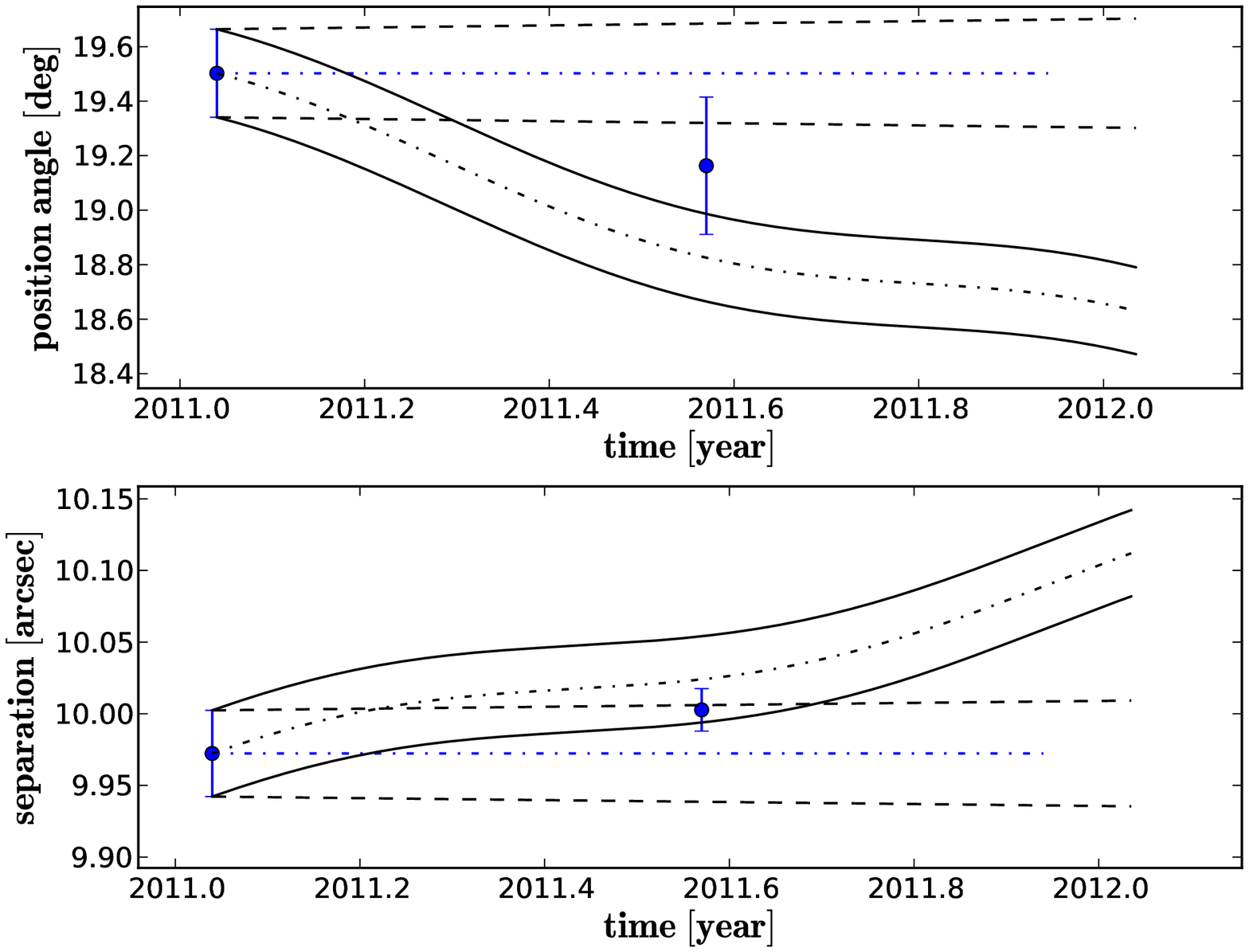}
\label{hd13-pm}
}
\subfigure[Companion candidate to HD\,185269]{
\includegraphics[scale=0.4]{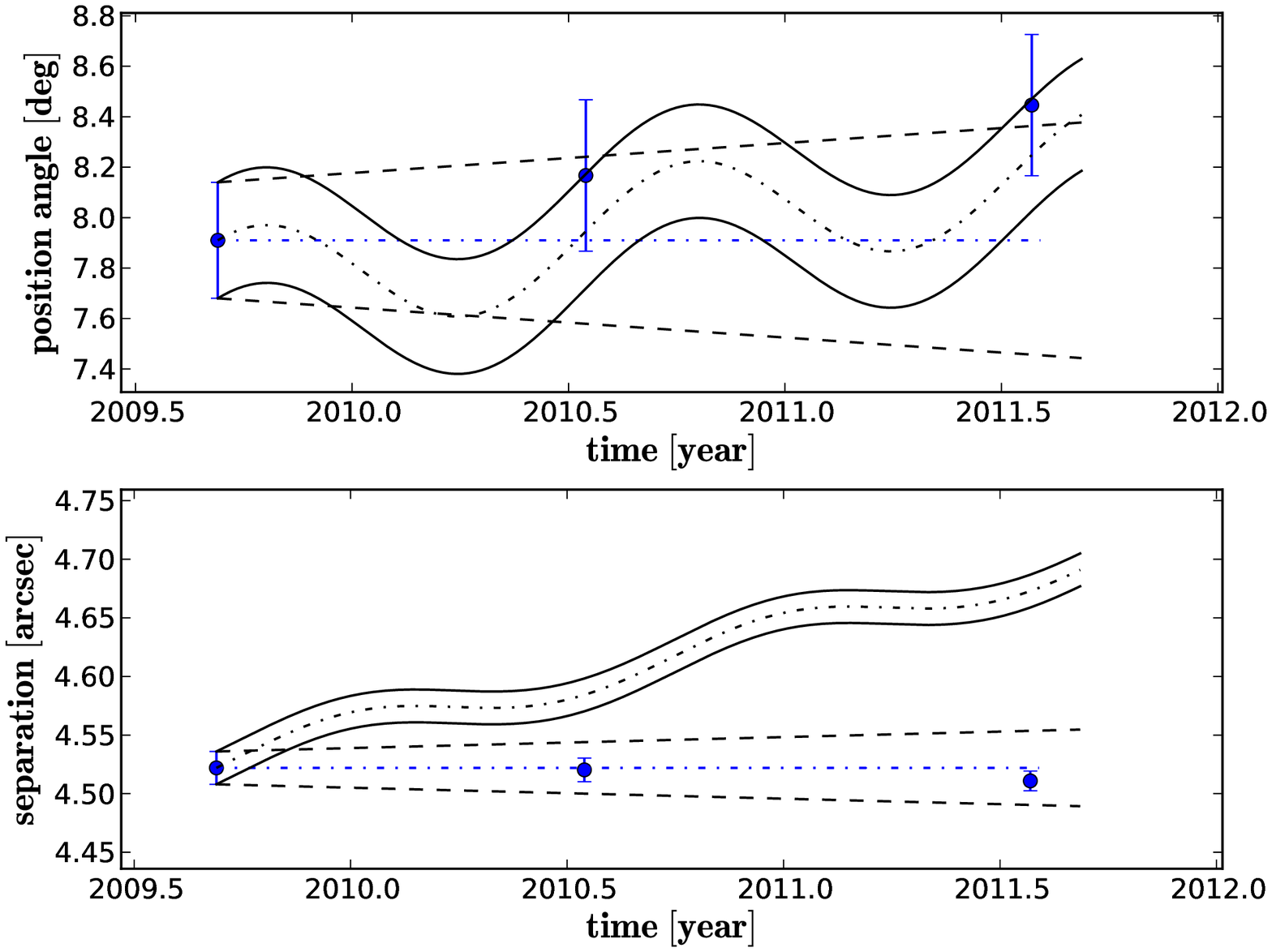}
\label{hd185-pm}
}
\subfigure[Companion candidate 1 to HD\,183263]{
\includegraphics[scale=0.4]{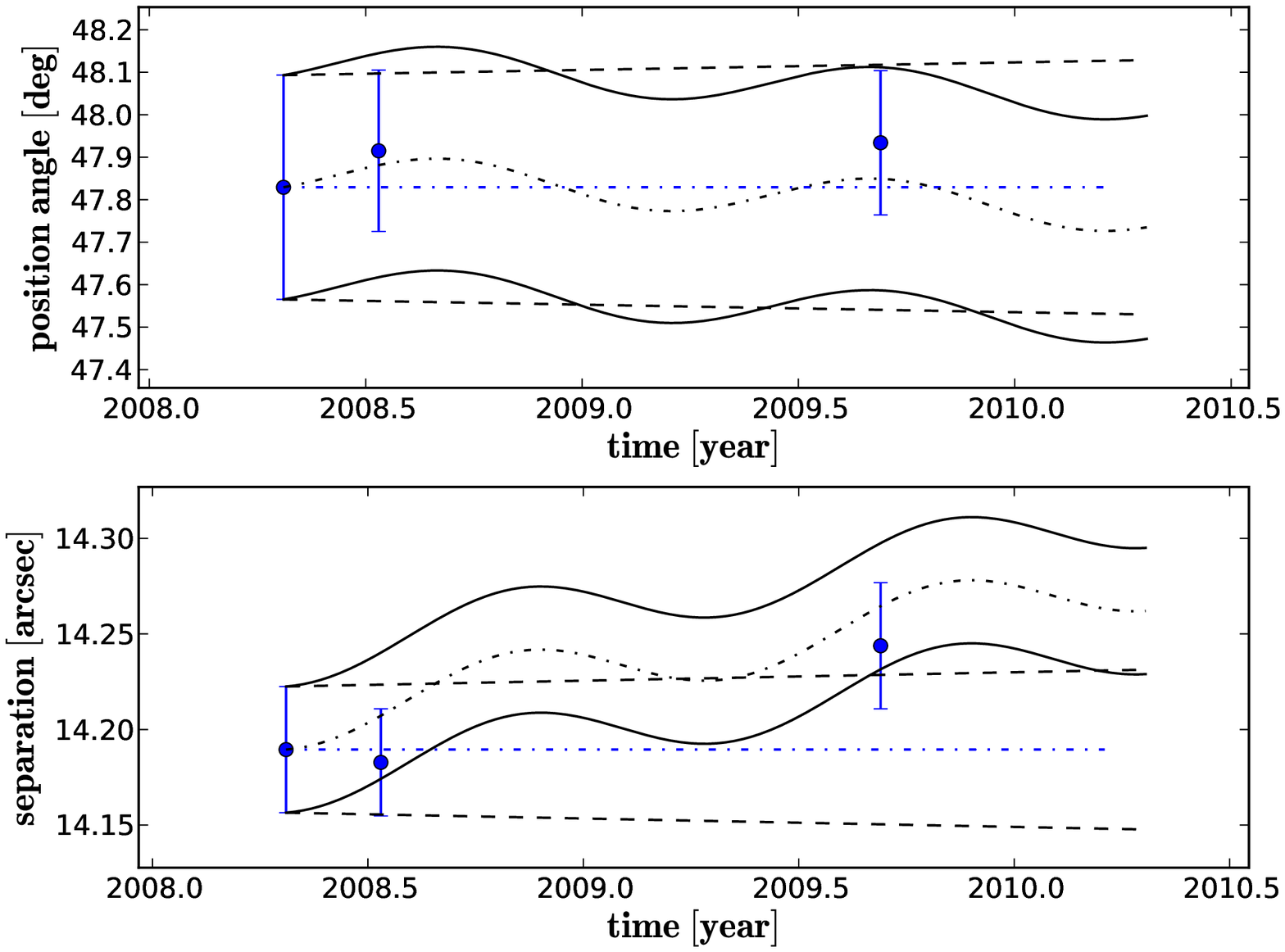}
\label{hd183-cc1-pm}
}
\subfigure[Companion candidate 2 to HD\,183263]{
\includegraphics[scale=0.4]{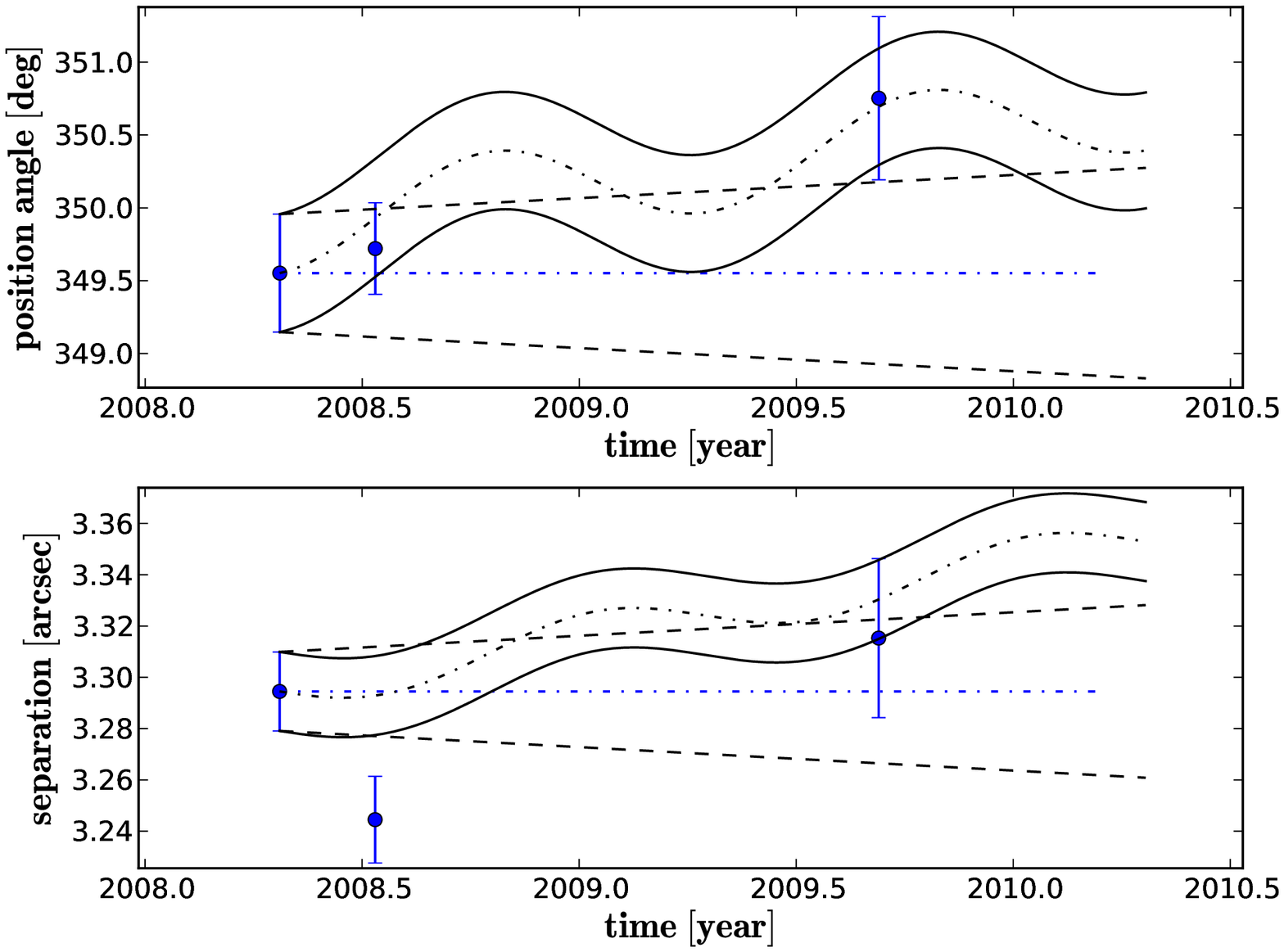}
\label{hd183-cc2-pm}
}
\subfigure[Companion candidate 1 to HD\,187123]{
\includegraphics[scale=0.4]{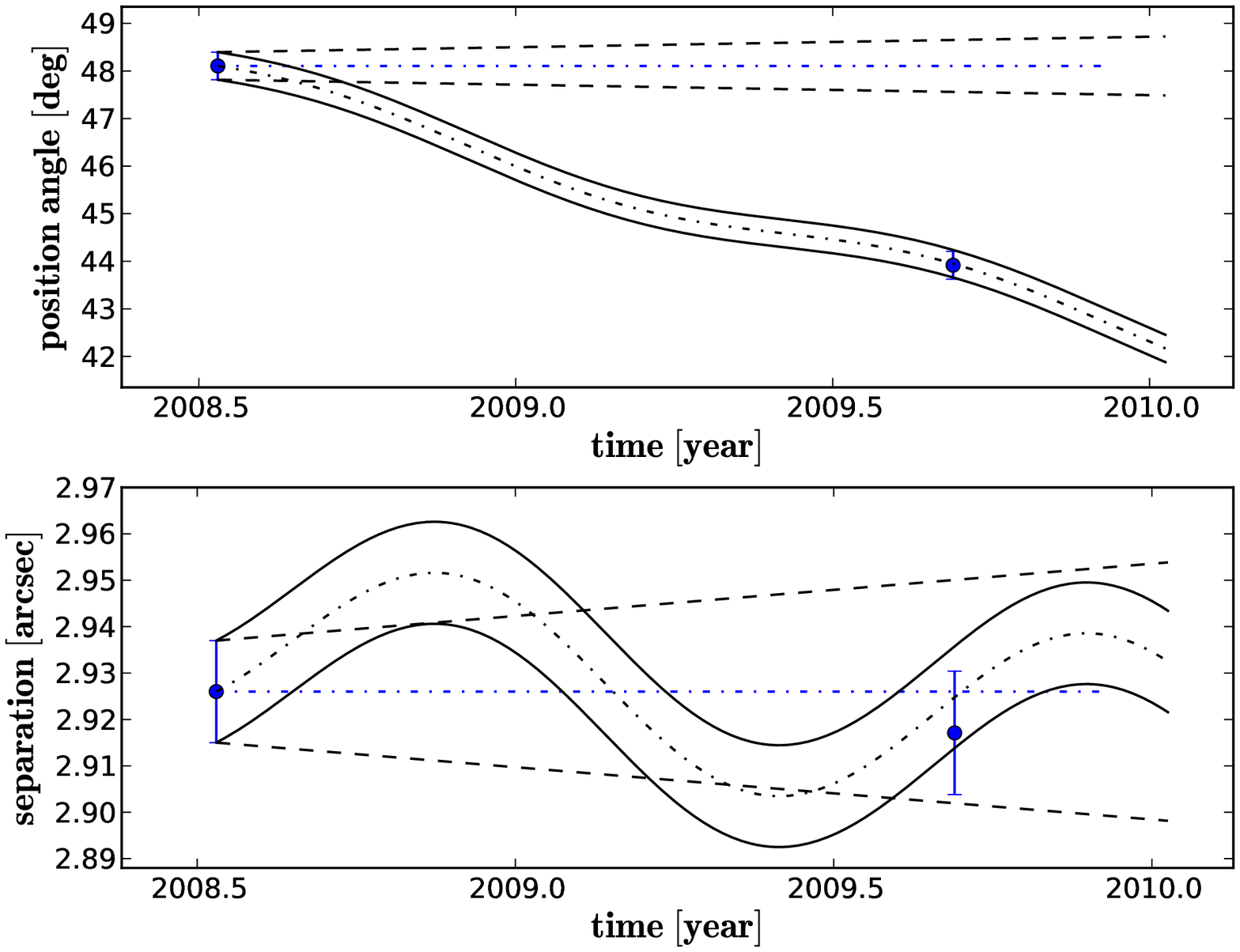}
\label{hd187-cc1-pm}
}
\subfigure[Companion candidate 2 to HD\,187123]{
\includegraphics[scale=0.4]{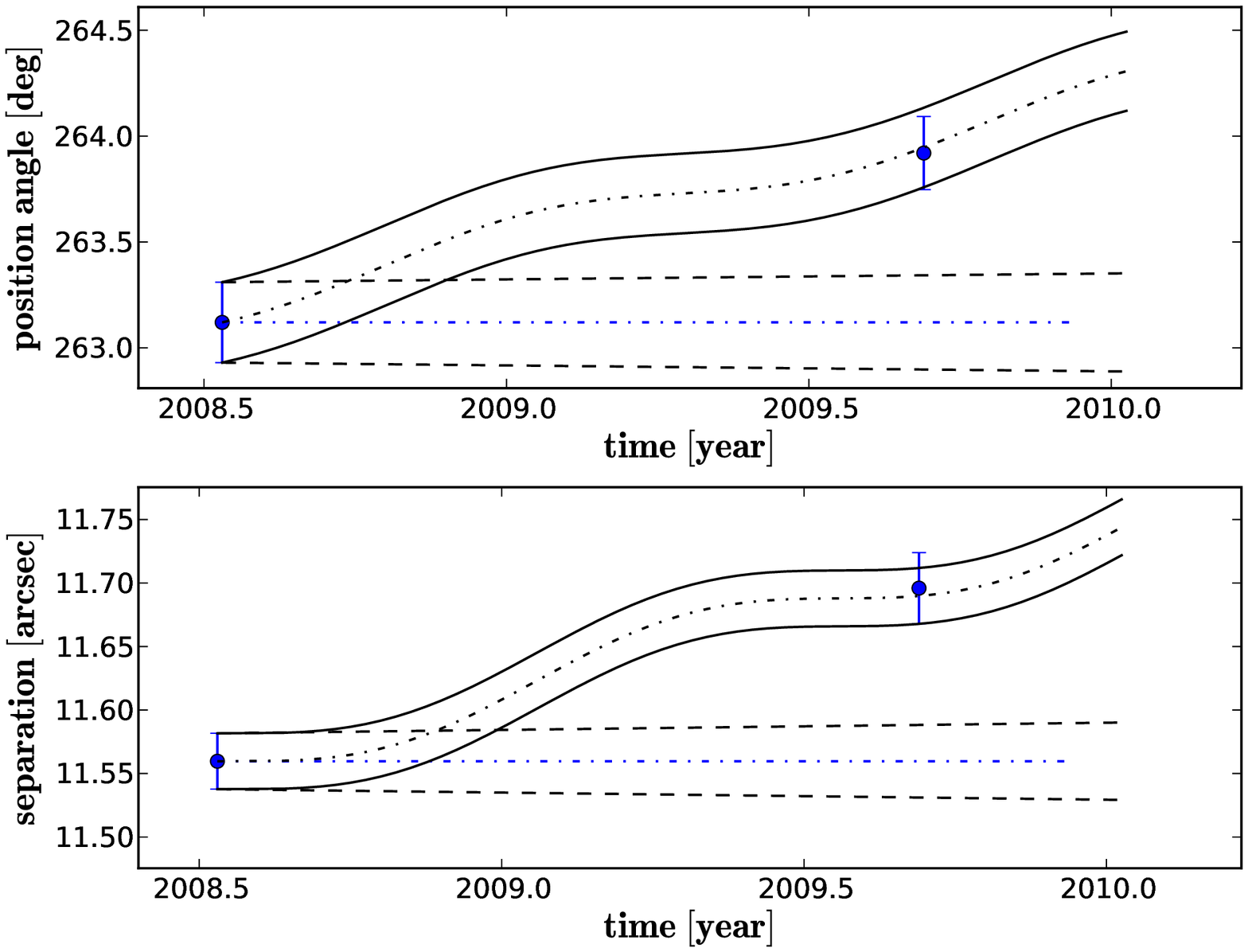}
\label{hd187-cc2-pm}
}
\caption[]{Proper motion diagrams for all newly discovered companion candidates. Upper diagrams are always for the position angle and lower diagrams for the separation, both plotted over time in years. The dashed lines mark the area for maximum possible orbital motion in case of bound objects and circular orbits. The solid lines mark the area for non-moving background objects. The wobble in the background hypothesis is introduced by the annual parallax.}
\label{fig:prop-mo-diag}
\end{figure*}

We performed a proper motion analysis for all companion candidates listed in Table~\ref{tab:cc-measurements} to distinguish between true physical companions and distant background objects. Hence we computed how separation and position angle would change from the first to the last observation epoch if the detected objects are indeed only in the background, and are therefore most likely standing still, also considering the effect of the annual parallax due to the earth's motion around the sun. We then compared these expected positions with the actual positions measured in our images. The resulting diagrams for all new companion candidates are listed in Fig.~\ref{fig:prop-mo-diag}. Even if the detected objects are physically associated to the primaries, they might show a slightly different proper motion, which can be explained by orbital motion around the primary star. To account for this, we also added to the analysis the maximum possible change of separation and position angle due to orbital motion at the projected distance of our first epoch measurement. We assumed an edge-on circular orbit for changes in separation and a face-on circular orbit for changes in position angle respectively. \\ 

\begin{figure}
\includegraphics[scale=0.355]{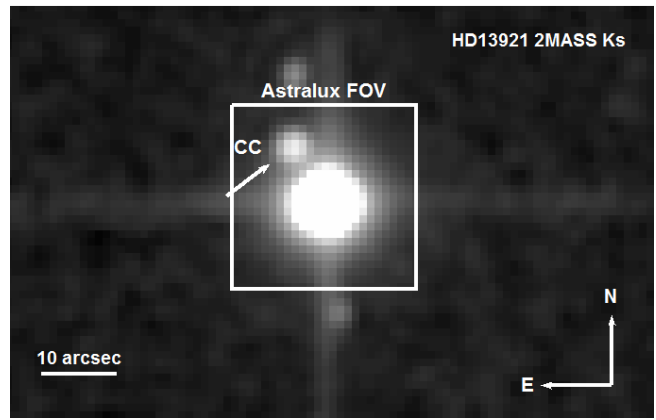}
\caption[]{2MASS Ks-Band Image of HD\,13931 in epoch 28/10/1998. The AstraLux field of view is indicated by the white rectangle.}
\label{fig:hd13931-2mass}
\end{figure}

\begin{figure}
\includegraphics[scale=0.4]{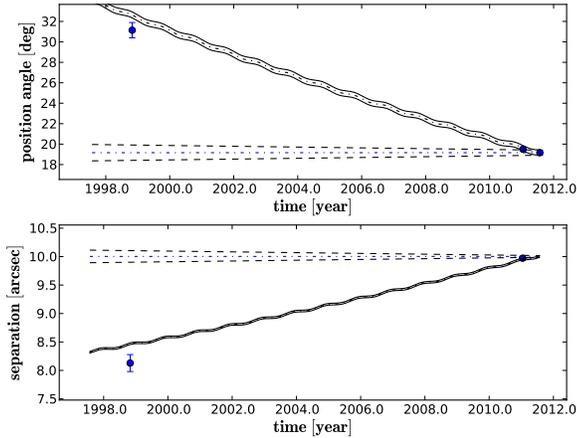}
\caption[]{Proper motion diagram for the companion candidate to HD\,13931 including a 2MASS measurement of epoch 28/10/1998. The areas for maximum orbital motion and background hypothesis are plotted backwards from our latest measurement in epoch 27/07/2011.}
\label{fig:hd13-pm-2mass}
\end{figure}

\emph{HD\,13931}\\
\\
HD\,13931 was observed in January and July 2011. The associated proper motion plot can be found in Fig.~\ref{hd13-pm}. The data point taken in July 2011 is, within its uncertainty, consistent with the background and common proper motion hypothesis. The common proper motion hypothesis can be rejected only on the $0.83$ and $1.09 \, \sigma$ level for separation and position angle respectively, hence it is not possible to draw a final conclusion from our AstraLux measurements.\\
This companion candidate was resolved in 2MASS (Two Micron All Sky Survey) J, H and Ks-band images taken on October 28th 1998. We show the corresponding Ks-band image in Fig.~\ref{fig:hd13931-2mass}. Our measurements for this epoch yield a separation of $8.13 \pm 0.15\,$arcsec and a position angle of $31.14^\circ \pm 0.75^\circ$. Using this new data point, we created a proper motion plot as shown in Fig.~\ref{fig:hd13-pm-2mass}. Although the 2MASS data points are not consistent with the background hypothesis within $1\,\sigma$, we can still reject common proper motion with $10.38\,\sigma$ for separation and $11.29\,\sigma$ for position angle. The inconsistency with the background hypothesis is most likely caused by a slow proper motion of the detected object. We conclude that our companion candidate around HD\,13931 is only a background object.\\

\pagebreak

\emph{HD\,185269}\\
\\
In Fig.~\ref{hd185-pm} we show the proper motion plot for the companion candidate next to HD\,185269. Due to the overlap of background and common proper motion area in the position angle plot, our angular measurements would be consistent with both hypothesis. However, our measurements of the separation are only consistent with an object of common proper motion with the primary star. We can reject the background hypothesis with $9.9 \, \sigma$, and are therefore concluding that our object is indeed a previously unknown stellar companion to HD\,185269. HD\,185269\,B has a projected separation of $227\,AU$. Further follow-up observations will be executed to determine the orbital motion of the B component.\\
Using IRAF (Image Reduction and Analysis Facility) standard aperture photometry, we calculate an average magnitude difference of $\Delta I \, = \, 7.2 \pm 0.2 \,$mag throughout all our observation epochs. Given the age of the primary of $4.2 \,$Gyr (\citealt{b20}) and the parallax of 21.11\,$\pm$\,0.74\,mas (\citealt{b2}), we find a mass of $0.239 \, \pm \, 0.022 \, M_{\odot}$, using the models by \cite{b9}.\\

\begin{figure}
\includegraphics[scale=0.355]{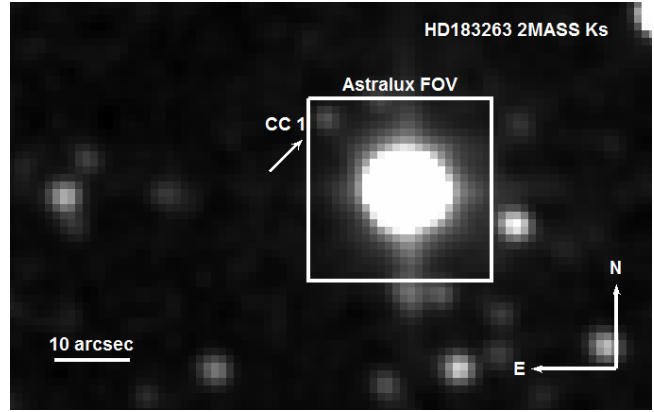}
\caption[]{2MASS Ks-Band Image of HD\,183263 in epoch 05/08/1999. The AstraLux field of view is indicated by the white rectangle.}
\label{fig:hd183-2mass}
\end{figure}

\begin{figure}
\includegraphics[scale=0.4]{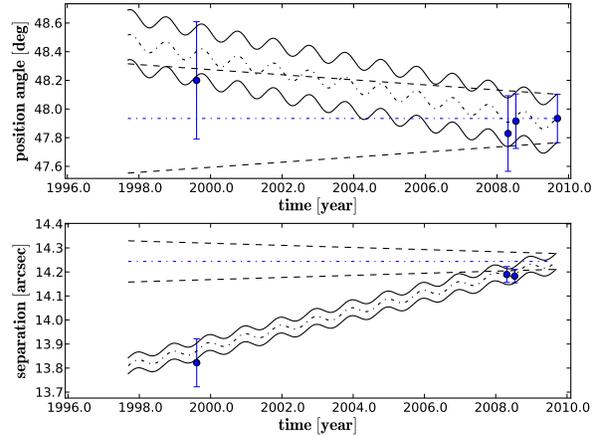}
\caption[]{Proper motion diagram for the first (far) companion candidate to HD\,183263 including a 2MASS measurement of epoch 08/1999. The areas for maximum orbital motion and background hypothesis are plotted backwards from our latest measurement in epoch 07/09/2009.}
\label{fig:pm-2mass}
\end{figure}

\emph{HD\,183263}\\
\\
Fig.~\ref{hd183-cc1-pm} and \ref{hd183-cc2-pm} show the proper motion plots for the two companion candidates discovered around HD\,183263.\\ 
Due to the comparatively small proper motion of only $38.09\,$mas$\cdot$yr$^{-1}$ ($\approx \, 0.8 \,$AstraLux pix), a final conclusion for the first companion candidate is difficult with our epoch difference of $1.25 \,$yr. The position angle gives no useful information, since the background and common proper motion area are completely overlapping, and we can only reject the common proper motion hypothesis with $1.06 \, \sigma$, in the case of the separation. We could, however, identify our first companion candidate once more in the 2MASS catalogue. We show the corresponding image in Ks-band in Fig.~\ref{fig:hd183-2mass}. The image was taken on August 5th 1999, giving us $10\,$yr of epoch difference. The 2MASS-PSC (2MASS - Point Source Catalogue, \citealt{b28}) measurements yield a separation of $13.8 \pm 0.1 \,$arcsec and a position angle of $48.2^{\circ} \pm 0.4^{\circ}$. Using this information, we created a new proper motion plot incorporating the 2MASS epoch which is shown in Fig.~\ref{fig:pm-2mass}. Even though the error bars are significantly larger than the ones of our AstraLux measurements, the new proper motion analysis in separation clearly indicates that the first companion candidate around HD\,183263 is a background object, with a significance of $3.3 \, \sigma$.\\
The second (closer) companion candidate was also detected in all three observation epochs. In July 2008 it was only barely visible, making these data points somewhat dubious. While the separation measurements are in general still consistent with a co-moving object as well as a background object, the development of the position angle indicates (with $1.4 \, \sigma$ corresponding to $83.8\,$\% probability) that the second companion candidate is a background object as well.\\ 

\pagebreak

\emph{HD\,187123}\\
\\
The proper motion plots for the two companion candidates around HD\,187123 are shown in Fig.~\ref{hd187-cc1-pm} and \ref{hd187-cc2-pm}.\\
Both companion candidates show as fully consistent with the background hypothesis with an average significance level of $3.3\, \sigma$. We conclude that both objects are indeed distant, non-moving background objects. HD\,187123 is a single star within the given detection limits described in section \ref{dl}. \\ 

\begin{figure}
\includegraphics[scale=0.4]{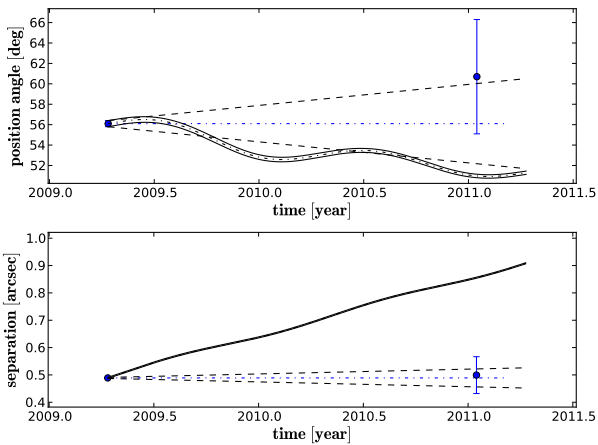}
\caption[]{Proper motion diagrams for the stellar companion to HD\,126614}
\label{fig:pm-hd126}
\end{figure}

\emph{HD\,126614}\\
\\
We also performed a proper motion analysis for the companion to HD\,126614, using the measurements by \cite{b17} as first epoch, and our own AstraLux measurement as second epoch. The resulting plot is shown in Fig.~\ref{fig:pm-hd126}.

Our measured separation of $499 \pm 67 \,$mas and position angle of $60.7^{\circ} \pm 5.6^{\circ}$ are consistent with the the astrometry of
the companion in the discovery epoch ($489.0 \pm 1.9 \,$mas and $56.1^{\circ} \pm 0.3^{\circ}$, as given by \citealt{b17}). However, due to the very low declination of $-05^\circ \, 10' \, 4''$, we had to observe HD\,126614 at high air mass, which led to an extended PSF in north-south direction. This made it very difficult to determine the position angle precisely, which is also evident in the large error bars of the measurement. Still the position angle is, within its error bars, consistent with common proper motion.\\
Despite the uncertainties in the position angle, we can still reject the background hypothesis with $1.7 \, \sigma$ and $5.3 \, \sigma$ for position angle and separation respectively. We can therefore conclude that HD\,126614\,B is indeed physically associated with A.\\
Due to the small separation between primary and companion, aperture photometry with IRAF is not viable. We could, however, infer the I-band magnitude difference of companion and primary from the ratio of the peak counts. We calculate $\Delta I \, = \, 5.59 \pm 0.15 \,$mag. Given the log(age)\,=\,9.1 (\citealt{b17}) and parallax of 14.63\,$\pm$\,1.15\,mass (\citealt{b2}), we find a mass of 0.307\,$\pm$\,0.033\,M$_{\odot}$, using the models by \cite{b9}. This is consistent with the mass of 0.324\,$\pm$\,0.004 found by \cite{b17}.

\subsection{Detection limits and non-detections}
\label{dl}

\begin{figure*}
\subfigure[average dynamic range in epoch 23/04/2008]{
\includegraphics[scale=0.33]{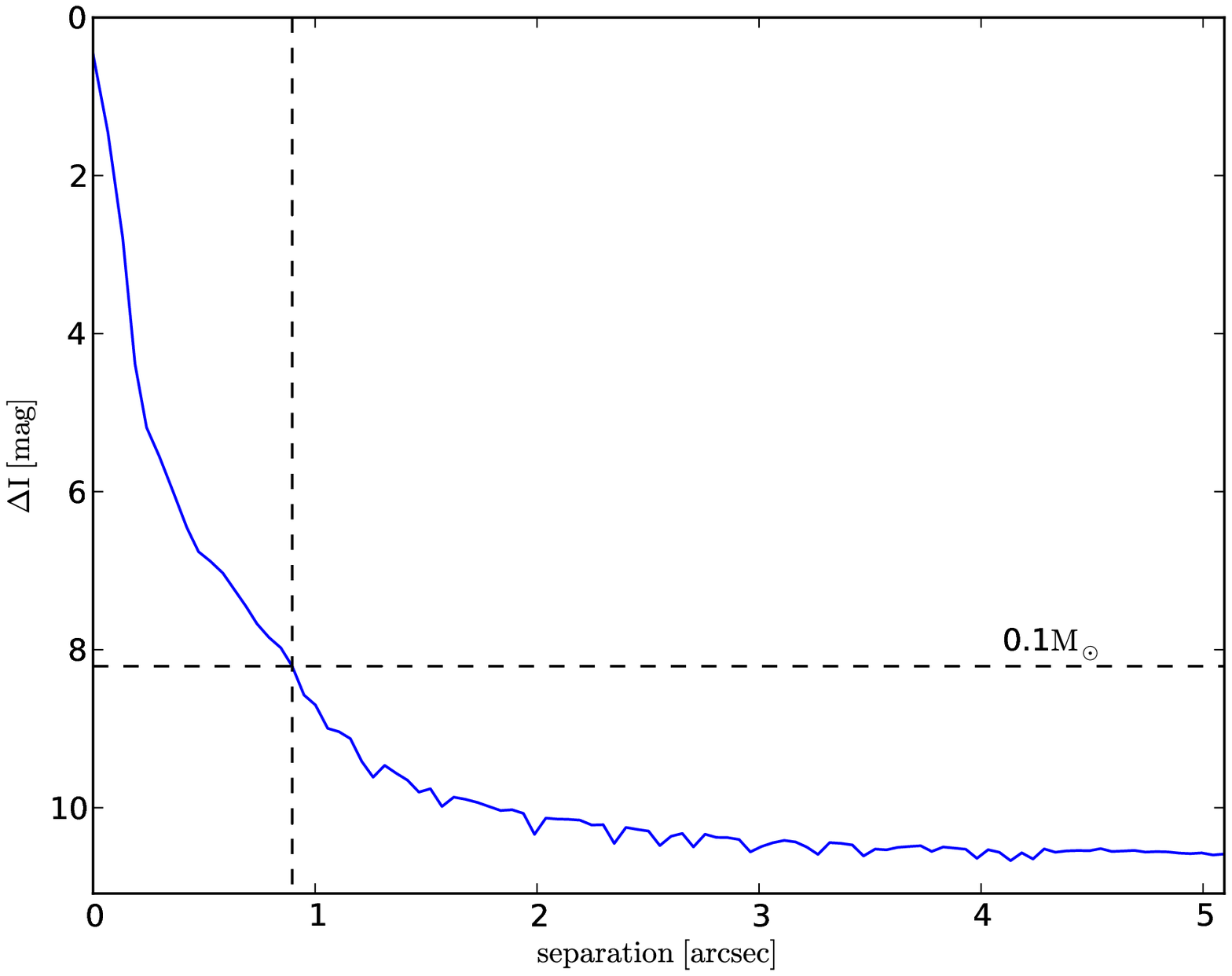}
\label{0408-dr}
}
\subfigure[average dynamic range in epoch 11/07/2008]{
\includegraphics[scale=0.33]{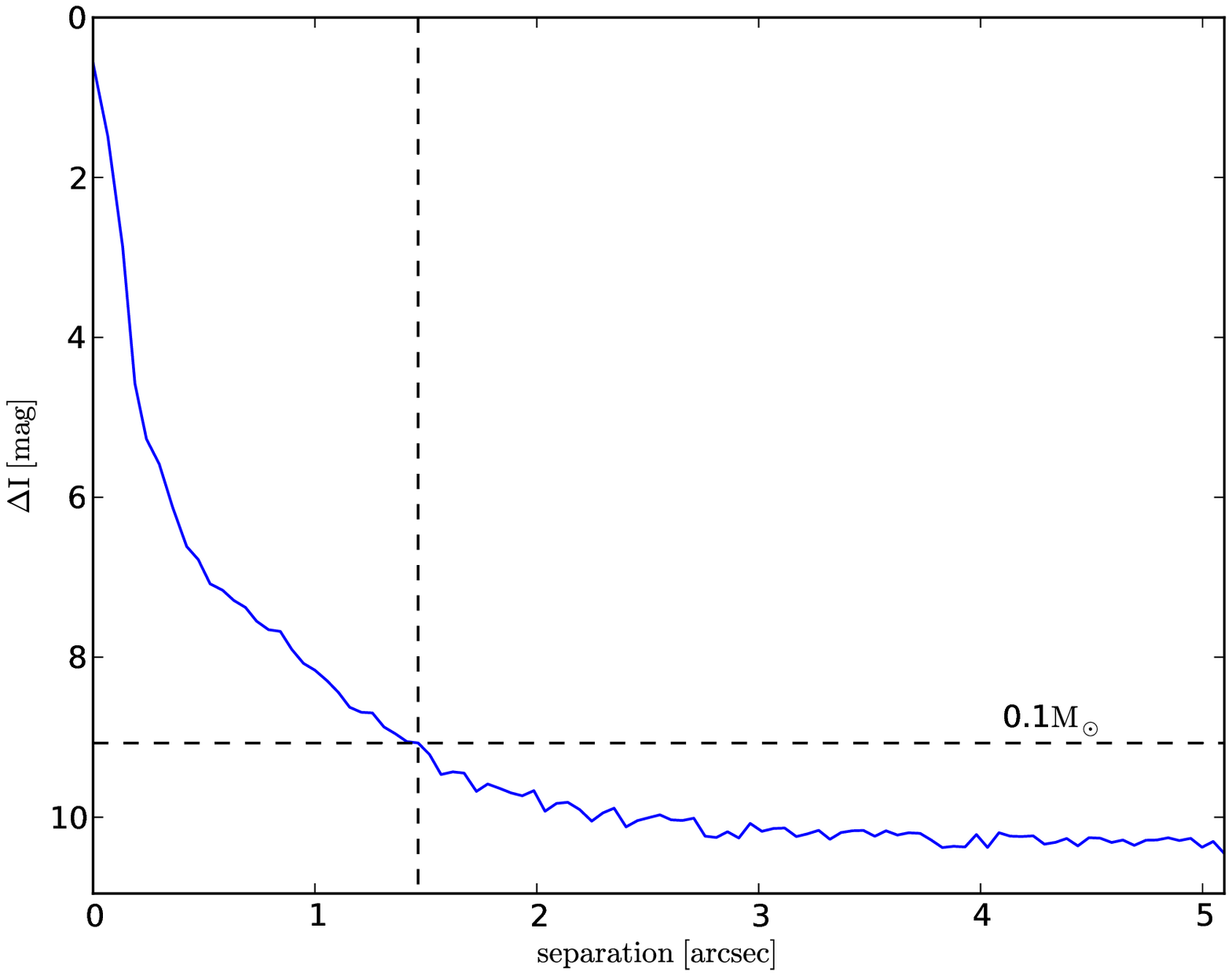}
\label{0708-dr}
}
\subfigure[average dynamic range in epoch 16/01/2009]{
\includegraphics[scale=0.33]{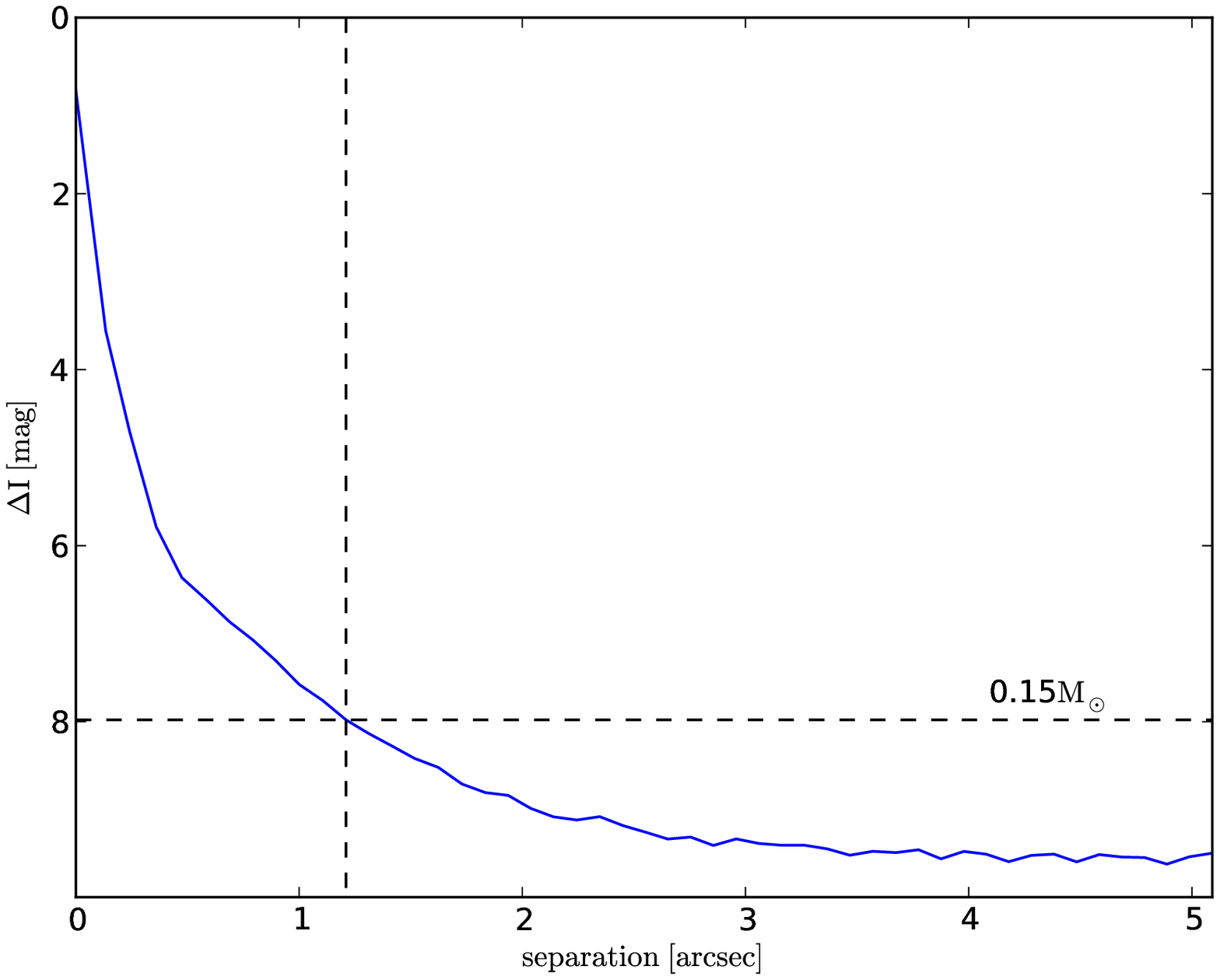}
\label{0109-dr}
}
\subfigure[average dynamic range in epoch 07/09/2009]{
\includegraphics[scale=0.33]{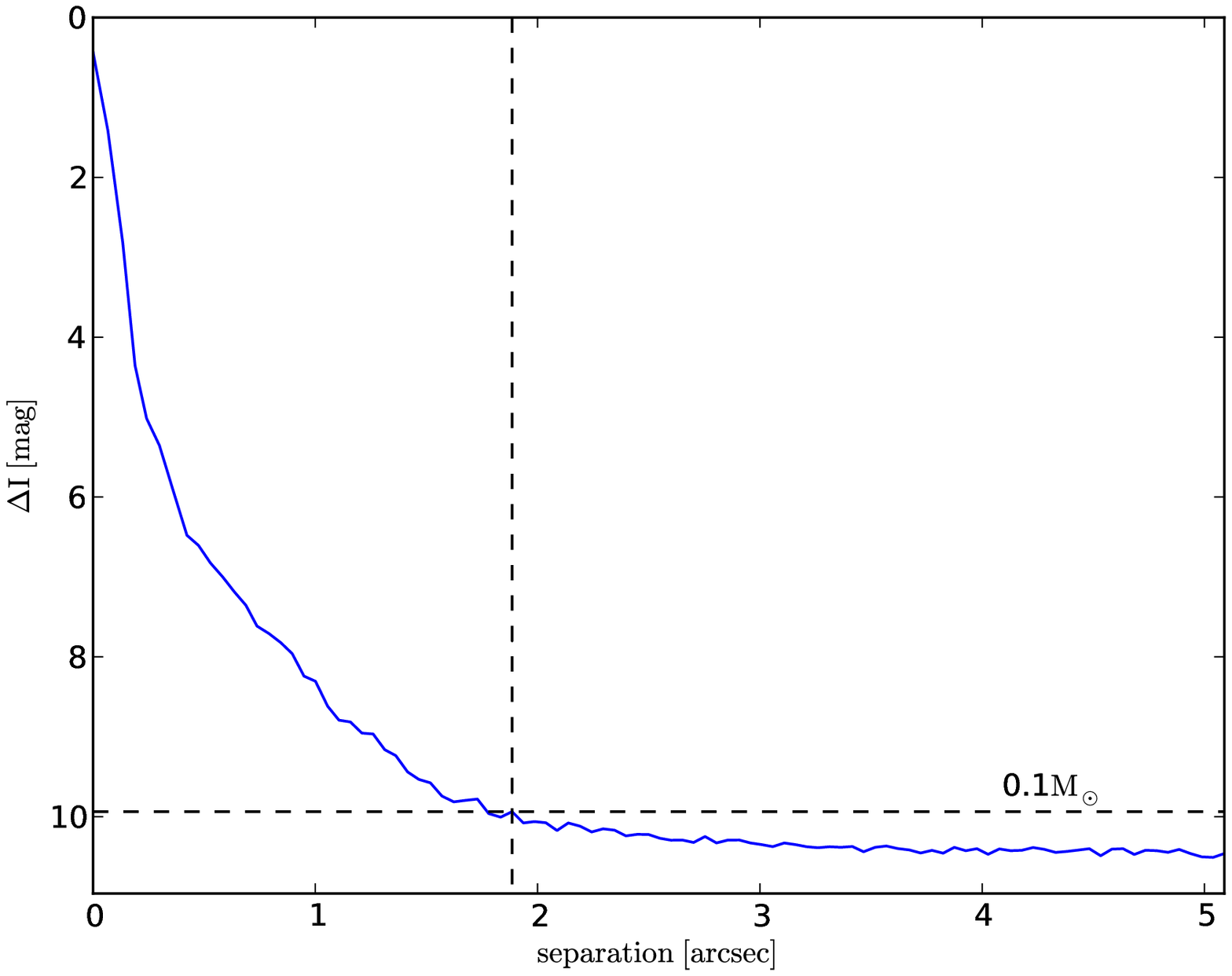}
\label{0909-dr}
}
\subfigure[average dynamic range in epoch 23/02/2010]{
\includegraphics[scale=0.33]{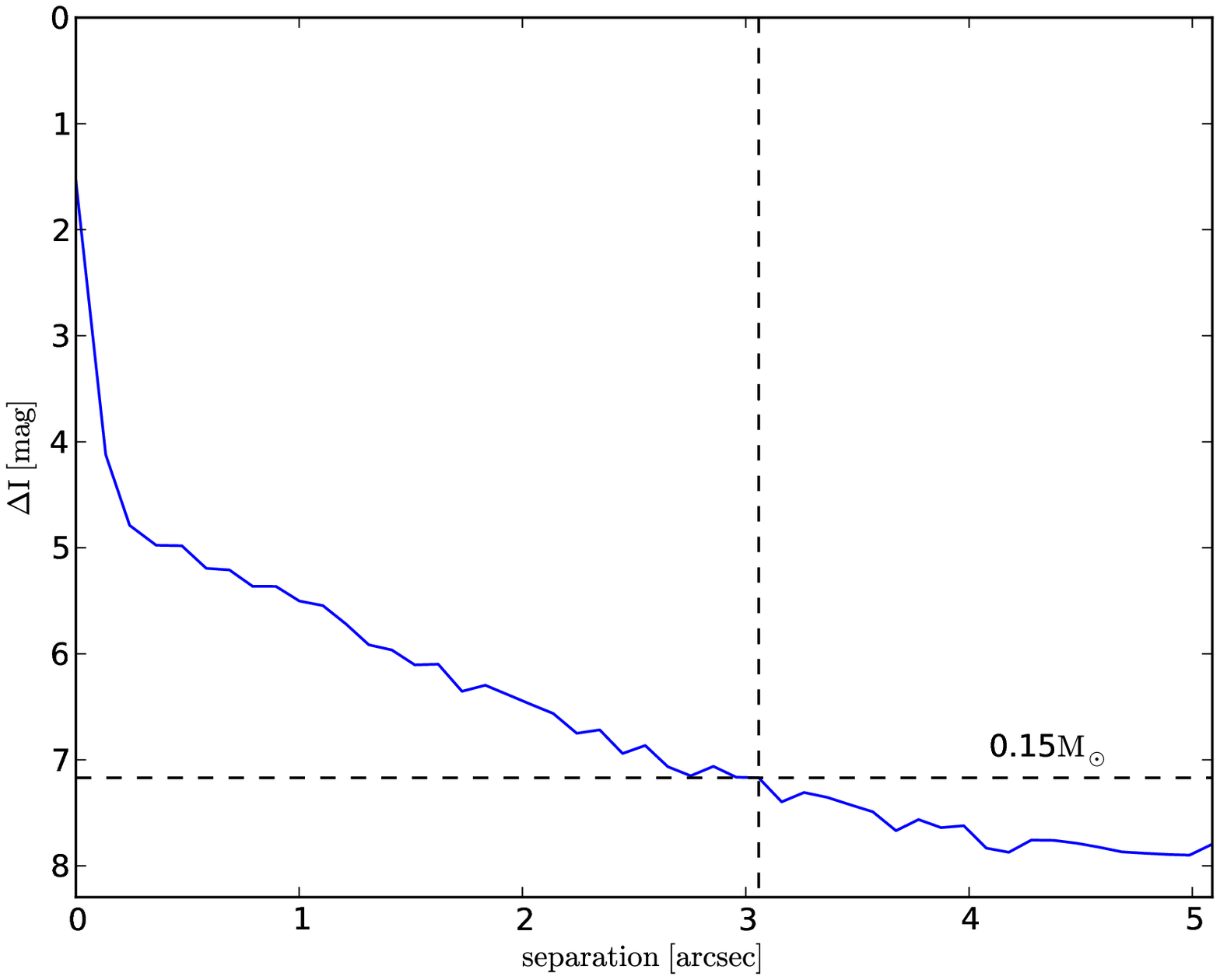}
\label{0210-dr}
}
\subfigure[average dynamic range in epoch 14/07/2010]{
\includegraphics[scale=0.33]{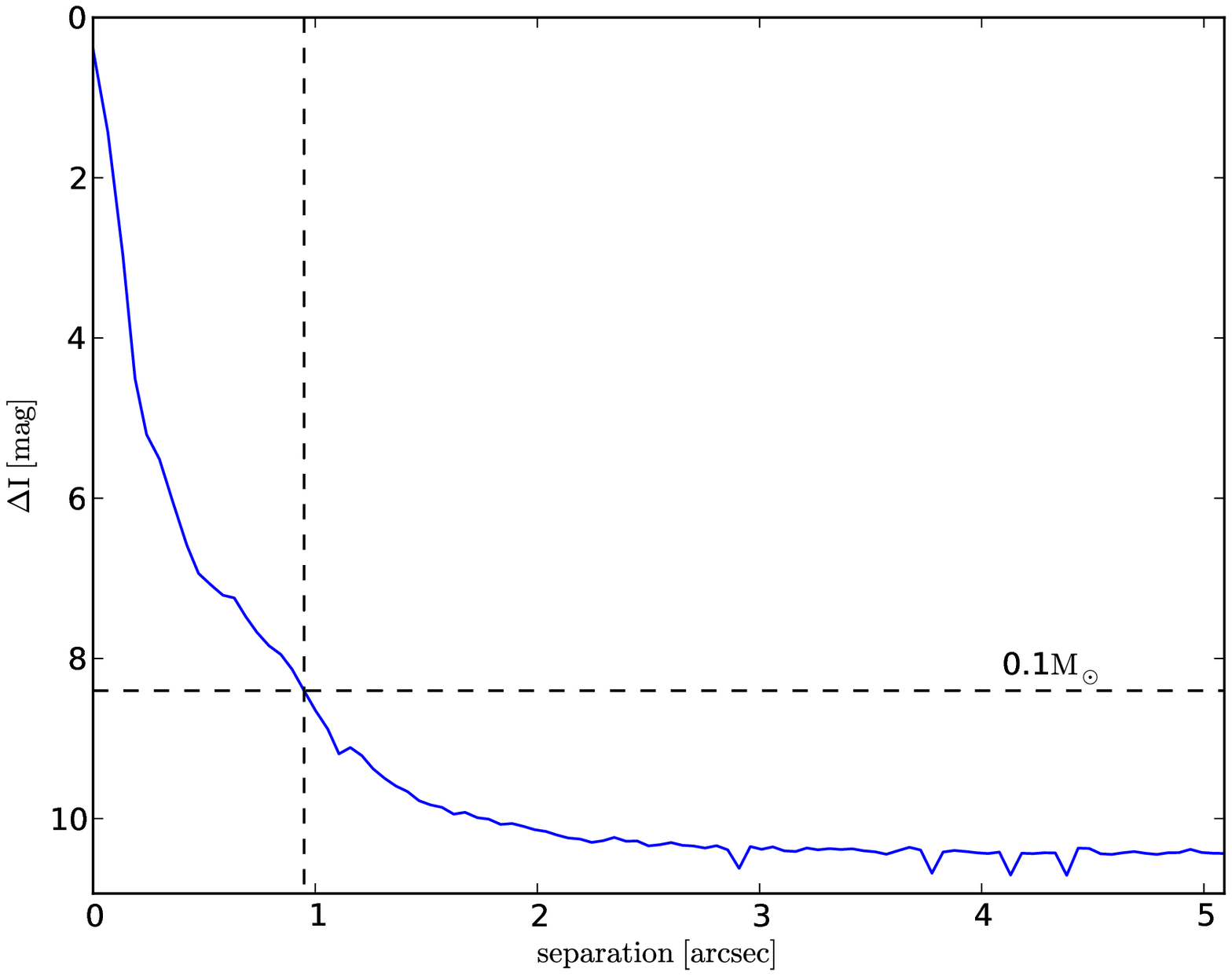}
\label{0710-dr}
}
\subfigure[average dynamic range in epoch 14/01/2011]{
\includegraphics[scale=0.33]{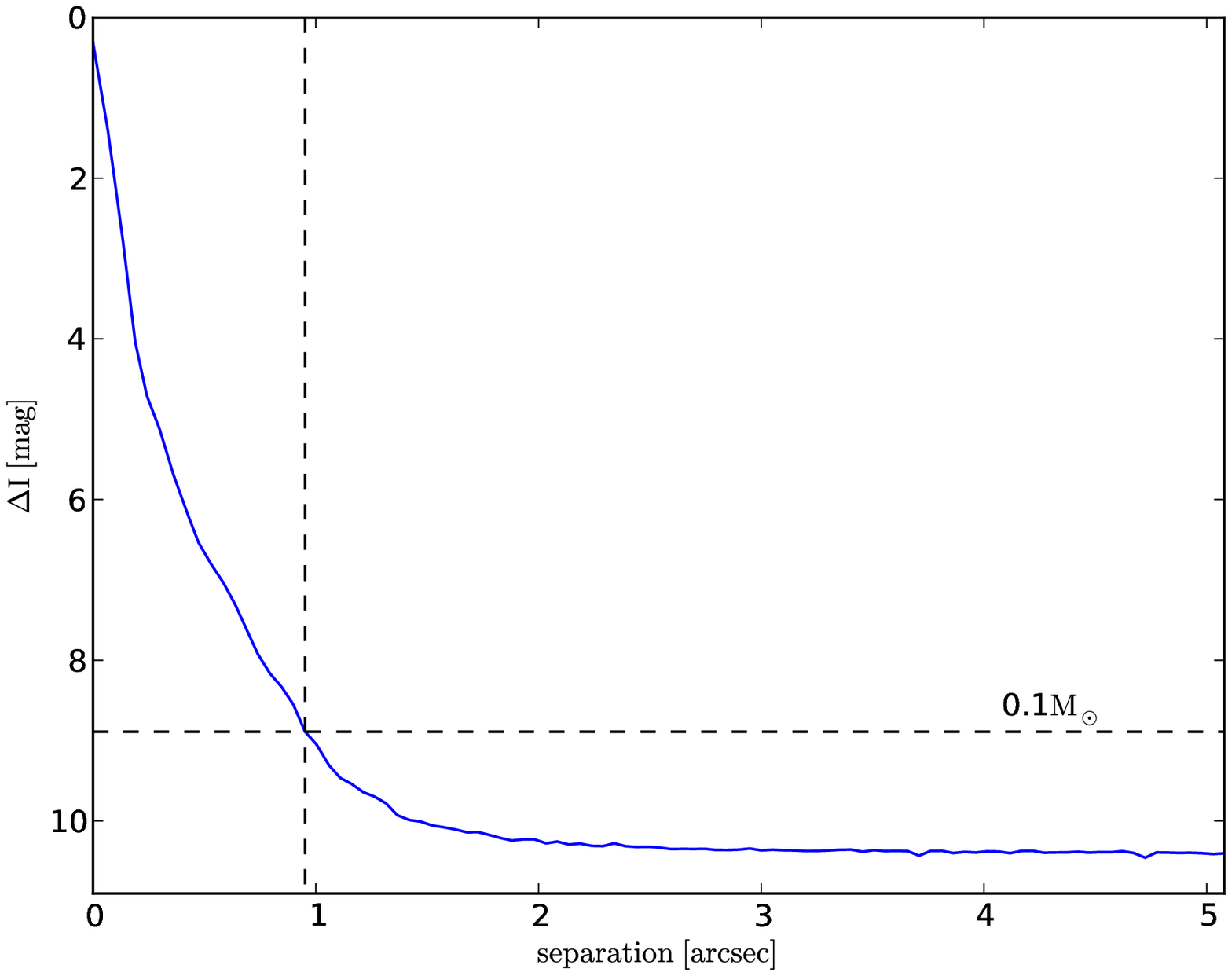}
\label{0111-dr}
}
\subfigure[average dynamic range in epoch 27/07/2011]{
\includegraphics[scale=0.33]{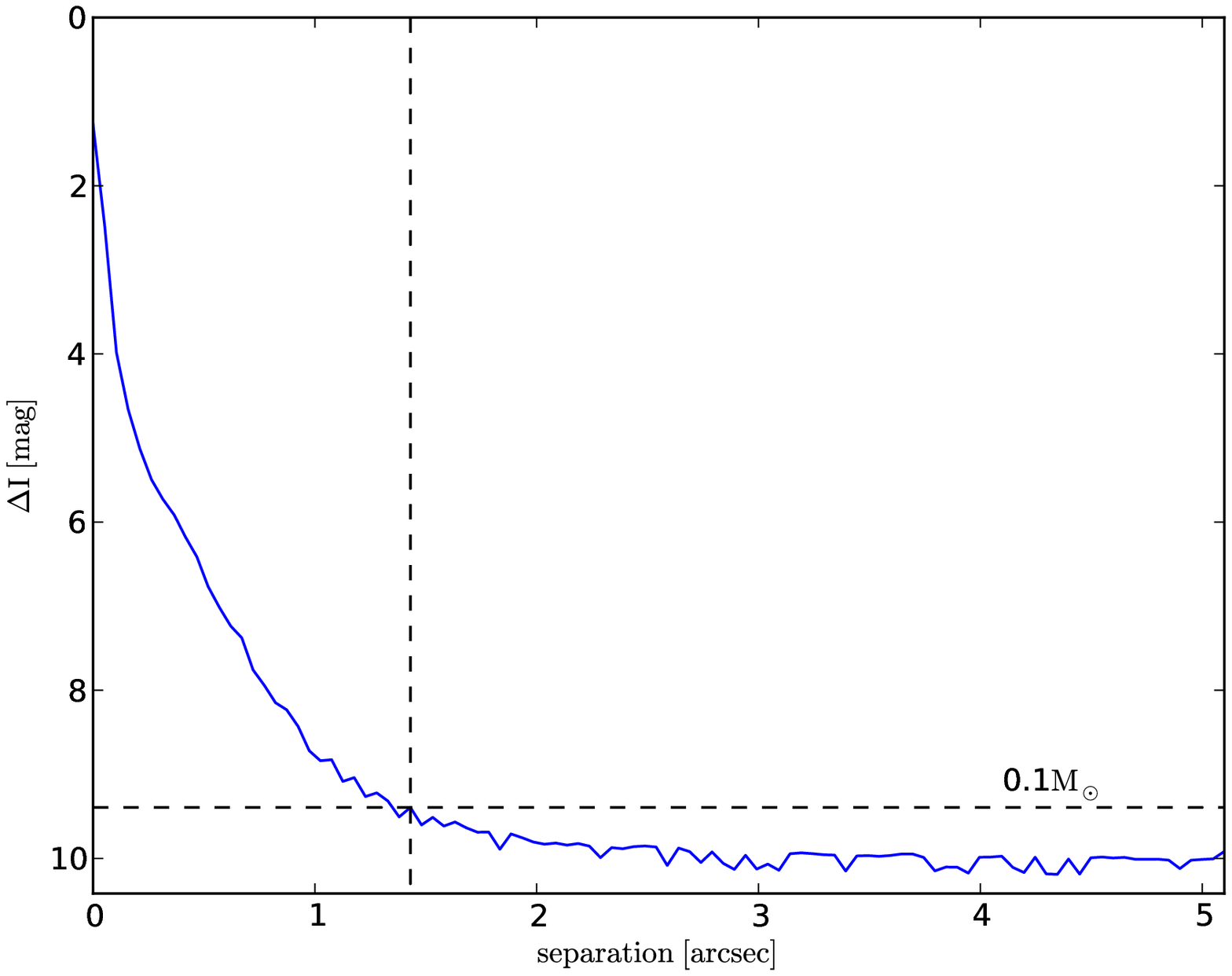}
\label{0711-dr}
}

\caption[]{Average dynamic range plots for a signal-to-noise of $5$ for all our observation epochs. The dashed lines represent the detectable minimum mass objects at a given separation from the primary star, using the model by \cite{b9}. The primary stars' PSF was always subtracted by unsharp mask filtering before determining the dynamic range.} 
\label{fig:dr}
\end{figure*}

In Table~\ref{tab:dl-ccs} and Table~\ref{tab:non-d} we list all the systems that were observed in our survey to date, with and without additional stellar components detected, respectively. We also list the epochs in which these systems have been observed, since our detection limits vary with the weather and seeing conditions. Additionally we present limits for the lowest mass objects detectable at separations of $0.5$, $1$, $2$ and $5\,$arcsec, at a signal-to-noise ratio of $5$, in the PSF subtracted images. We calculated the uncertainties of these mass limits considering the uncertainties in the measured parallaxes for each system, as well as a maximum photometric error of 0.2\,mag in our Astralux images, along with an average photometric error of 0.05\,mag in the primary stars' I-band magnitudes.\\
In Fig.~\ref{fig:dr} we show the average dynamic range plots for each of our observation epochs. We determined the average dynamic range by measuring the noise levels in all reduced images of a given epoch with a frame selection rate of $5 \, \%$, assuming a signal-to-noise ratio of $5$. The primary stars' PSF was always subtracted to get the best possible results.\\
In all but two epochs we reach a magnitude difference of $\Delta I \, = \, 10 \,$mag outside of $2 \,$arcsec from the primary star. Given the average absolute magnitudes and ages of the target stars in each epoch as listed in Table~\ref{tab:dr-average} (calculated from individual values as listed in Simbad and exoplanet.eu respectively), and using the models by \cite{b9}, we would have easily been able to detect all wide stellar companions. Furthermore, we still probe the low mass stellar regime ($M \, < \, 0.15 M_{\odot}$) down to separations of $1 \,$arcsec from the primary star.\\
The observation epochs in January 2009 and February 2010 suffered from bad weather conditions, and while we could complete most of our observation program in January 2009, we could only observe for approximately one hour through the cloud cover in February 2010. This is reflected in the dynamic range graphs \ref{0109-dr} and \ref{0210-dr}, which show a noticeable decline in sensitivity in comparison with the other observation epochs. On average we are still able to detect stellar companions down to $0.15\,M_{\odot}$ outside of $1.2 \,$arcsec and $3.1 \,$arcsec respectively.\\
If we average all minimum masses for the respective separations as derived in Table~\ref{tab:non-d}, we conservatively get detection limits of $0.34 \, M_{\odot}$ at $0.5\,$arcsec as well as $0.22\, M_{\odot}$ at $1\,$arcsec and $0.15 \, M_{\odot}$ outside of $2\,$arcsec. Given our field of view, we can detect such low-mass stellar companions up to separations of $12\,$arcsec.

\begin{table}
  \caption{Average absolute magnitudes and ages of all target stars in an observation epoch}
  \begin{tabular}{@{}ccc@{}}
  \hline
        
 Epoch 		& MI [mag]		& Age [Gyr]	\\
 \hline
 23/04/2008 	& $4.32$ 				& $4.9$				\\
 11/07/2008 	& $3.43$ 				& $6.0$				\\
 16/01/2009 	& $2.93$ 				& $4.6$				\\
 07/09/2009 	& $2.64$ 				& $6.2$				\\
 23/02/2010 	& $3.72$ 				& $2.2$				\\
 14/07/2010 	& $4.14$ 				& $7.4$				\\
 14/01/2011	& $3.67$ 				& $5.4$				\\
 27/07/2011 	& $3.20$ 				& $6.5$				\\						
\hline\end{tabular}
\label{tab:dr-average}
\end{table}

\begin{table*}
  \caption{List of all stars with companions and companion candidates detected. We give the numbers of frames observed, each frame with an exposure time of $29.54\,$ms. We also give an upper limit for the minimum mass detectable at $0.5$, $1$, $2$ and $5\,$arcsec using the models by \protect\cite{b9} and assuming an average age for each epoch as listed in Table~\ref{tab:dr-average}.}
  \label{tab:dl-ccs}
  \begin{threeparttable}
  \begin{tabular}{@{}lcccccc@{}}
  \hline
  Star			& Epoch	& \# of frames & M $[M_{\odot}]$ at $0.5''$	& M $[M_{\odot}]$ at $1''$ & M $[M_{\odot}]$ at $2''$ & M $[M_{\odot}]$ at $5''$	\\
  \hline
HD13931	&	14/01/2011	&	50000	&	0.173	$\pm$	0.014	&	0.097	$\pm$	0.003	&	0.086	$\pm$	0.001	&	0.085	$\pm$	0.001	\\
	&	27/07/2011	&	50000	&	0.365	$\pm$	0.036	&	0.142	$\pm$	0.010	&	0.105	$\pm$	0.004	&	0.100	$\pm$	0.004	\\
HD183263	&	23/04/2008	&	50000	&	0.194	$\pm$	0.020	&	0.099	$\pm$	0.003	&	0.086	$\pm$	0.001	&	0.085	$\pm$	0.001	\\
	&	11/07/2008	&	80000	&	0.180	$\pm$	0.016	&	0.144	$\pm$	0.011	&	0.099	$\pm$	0.003	&	0.088	$\pm$	0.002	\\
	&	07/09/2009	&	80000	&	0.178	$\pm$	0.015	&	0.096	$\pm$	0.003	&	0.087	$\pm$	0.001	&	0.086	$\pm$	0.002	\\
HD187123	&	11/07/2008	&	77800	&	0.149	$\pm$	0.010	&	0.103	$\pm$	0.004	&	0.086	$\pm$	0.001	&	0.083	$\pm$	0.001	\\
	&	07/09/2009	&	80000	&	0.160	$\pm$	0.012	&	0.091	$\pm$	0.002	&	0.084	$\pm$	0.001	&	0.084	$\pm$	0.001	\\
HD185269	&	07/09/2009	&	50000	&	0.307	$\pm$	0.032	&	0.166	$\pm$	0.013	&	0.107	$\pm$	0.005	&	0.096	$\pm$	0.003	\\
	&	14/07/2010	&	60000	&	0.286	$\pm$	0.027	&	0.125	$\pm$	0.007	&	0.093	$\pm$	0.003	&	0.091	$\pm$	0.002	\\
	&	27/07/2011	&	50000	&	0.239	$\pm$	0.021	&	0.113	$\pm$	0.006	&	0.091	$\pm$	0.002	&	0.089	$\pm$	0.002	\\
$\tau\,$Boo	&	23/04/2008	&	50000	&	0.208	$\pm$	0.019	&	0.104	$\pm$	0.004	&	0.088	$\pm$	0.001	&	0.088	$\pm$	0.001	\\
HD176051	&	27/07/2011	&	50000	&	0.149	$\pm$	0.010	&	0.095	$\pm$	0.003	&	0.084	$\pm$	0.001	&	0.084	$\pm$	0.001	\\
HD126614	&	14/01/2011	&	50000	&	0.179	$\pm$	0.018	&	0.103	$\pm$	0.005	&	0.090	$\pm$	0.002	&	0.088	$\pm$	0.002	\\

	\hline\end{tabular}
	\end{threeparttable}
\end{table*}

\begin{table*} \scriptsize
  \caption{List of all stars with no additional stellar components detected. We give the numbers of frames observed, each frame with an exposure time of $29.54\,$ms. We also give an upper limit for the minimum mass detectable at $0.5$, $1$, $2$ and $5\,$arcsec using the models by \protect\cite{b9} and assuming an average age for each epoch as listed in Table~\ref{tab:dr-average}.}
  \label{tab:non-d}
  \begin{threeparttable}
  \begin{tabular}{@{}lcccccccc@{}}
  \hline
  Star			& RA & DEC & Epoch	& \# of frames & M $[M_{\odot}]$ at $0.5''$	& M $[M_{\odot}]$ at $1''$ & M $[M_{\odot}]$ at $2''$ & M $[M_{\odot}]$ at $5''$	\\
  \hline
HD\,1461 	&	00 18 41.8	&	-08 03 10.8	&	14/07/2010	&	55000	&	0.157	$\pm$	0.011	&	0.099	$\pm$	0.003	&	0.083	$\pm$	0.001	&	0.082	$\pm$	0.001	\\
BD-1763 	&	00 28 34.3	&	-16 13 34.8	&	07/09/2009	&	50000	&	0.110	$\pm$	0.006	&	0.091	$\pm$	0.003	&	0.083	$\pm$	0.001	&	0.079	$\pm$	0.001	\\
Hat-P-19 	&	00 38 04.0	&	+34 42 41.6	&	27/07/2011	&	50000	&	0.140	$\pm$	0.022	&	0.109	$\pm$	0.012	&	0.099	$\pm$	0.008	&	0.098	$\pm$	0.007	\\
HD\,4203 	&	00 44 41.2	&	+20 26 56.1	&	11/07/2008	&	80000	&	0.176	$\pm$	0.020	&	0.117	$\pm$	0.008	&	0.091	$\pm$	0.003	&	0.086	$\pm$	0.002	\\
HD\,5319 	&	00 55 01.4	&	+00 47 22.4	&	11/07/2008	&	50000	&	0.424	$\pm$	0.047	&	0.221	$\pm$	0.027	&	0.140	$\pm$	0.013	&	0.128	$\pm$	0.011	\\
HIP\,5158 	&	01 06 02.0	&	-22 27 11.3	&	14/01/2011	&	50000	&	0.112	$\pm$	0.008	&	0.089	$\pm$	0.002	&	0.080	$\pm$	0.002	&	0.079	$\pm$	0.001	\\
HD\,6718 	&	01 07 48.6	&	-08 14 01.3	&	14/01/2011	&	50000	&	0.211	$\pm$	0.021	&	0.107	$\pm$	0.005	&	0.085	$\pm$	0.001	&	0.083	$\pm$	0.001	\\
HD\,7924 	&	01 21 59.1	&	+76 42 37.0	&	27/07/2011	&	50000	&	0.100	$\pm$	0.003	&	0.082	$\pm$	0.001	&	0.079	$\pm$	0.001	&	0.078	$\pm$	0.001	\\
HD\,8673 	&	01 26 08.7	&	+34 34 46.9	&	14/01/2011	&	50000	&	0.270	$\pm$	0.025	&	0.113	$\pm$	0.006	&	0.089	$\pm$	0.002	&	0.088	$\pm$	0.001	\\
HD\,9446 	&	01 33 20.1	&	+29 15 54.5	&	14/07/2010	&	55000	&	0.148	$\pm$	0.011	&	0.099	$\pm$	0.003	&	0.086	$\pm$	0.001	&	0.083	$\pm$	0.002	\\
HD\,16232 	&	02 37 00.5	&	+24 38 49.9	&	23/02/2010	&	20000	&	0.402	$\pm$	0.033	&	0.318	$\pm$	0.034	&	0.208	$\pm$	0.022	&	0.121	$\pm$	0.007	\\
HD\,16175 	&	02 37 01.9	&	+42 03 45.4	&	27/07/2011	&	50000	&	0.236	$\pm$	0.024	&	0.108	$\pm$	0.005	&	0.093	$\pm$	0.003	&	0.091	$\pm$	0.003	\\
HD\,16760 	&	02 42 21.3	&	+38 37 07.2	&	23/02/2010	&	20000	&	0.289	$\pm$	0.047	&	0.223	$\pm$	0.036	&	0.159	$\pm$	0.019	&	0.108	$\pm$	0.009	\\
HD\,17092 	&	02 46 22.1	&	+49 39 11.1	&	14/07/2010	&	55000	&	0.441	$\pm$	0.181	&	0.213	$\pm$	0.122	&	0.121	$\pm$	0.043	&	0.112	$\pm$	0.035	\\
HIP\,12961 	&	02 46 42.8	&	-23 05 11.8	&	14/01/2011	&	50000	&	0.105	$\pm$	0.004	&	0.077	$\pm$	0.001	&	0.074	$\pm$	0.001	&	0.074	$\pm$	0.001	\\
HD\,17156 	&	02 49 44.4	&	+71 45 11.6	&	23/02/2010	&	20000	&	0.465	$\pm$	0.038	&	0.376	$\pm$	0.038	&	0.243	$\pm$	0.027	&	0.136	$\pm$	0.010	\\
HD\,22781 	&	03 40 49.5	&	+31 49 34.6	&	27/07/2011	&	50000	&	0.112	$\pm$	0.005	&	0.086	$\pm$	0.001	&	0.079	$\pm$	0.001	&	0.078	$\pm$	0.001	\\
HD\,28305\tnote{1} 	&	04 28 36.9	&	+19 10 49.5	&	07/09/2009	&	10000	&	0.862	$\pm$	0.031	&	0.653	$\pm$	0.028	&	0.342	$\pm$	0.036	&	0.325	$\pm$	0.034	\\
HD\,32518 	&	05 09 36.7	&	+69 38 21.8	&	14/01/2011	&	50000	&	0.686	$\pm$	0.036	&	0.259	$\pm$	0.030	&	0.183	$\pm$	0.020	&	0.185	$\pm$	0.020	\\
HD\,34445 	&	05 17 40.9	&	+07 21 12.0	&	14/01/2011	&	50000	&	0.219	$\pm$	0.020	&	0.094	$\pm$	0.003	&	0.089	$\pm$	0.002	&	0.089	$\pm$	0.002	\\
HD\,33564 	&	05 22 33.5	&	+79 13 52.1	&	07/09/2009	&	50000	&	0.253	$\pm$	0.022	&	0.148	$\pm$	0.009	&	0.097	$\pm$	0.003	&	0.095	$\pm$	0.002	\\
HD\,38801 	&	05 47 59.1	&	-08 19 39.7	&	14/01/2011	&	50000	&	0.363	$\pm$	0.065	&	0.115	$\pm$	0.012	&	0.097	$\pm$	0.006	&	0.095	$\pm$	0.005	\\
HD\,45652 	&	06 29 13.2	&	+10 56 02.0	&	16/01/2009	&	22610	&	0.165	$\pm$	0.013	&	0.118	$\pm$	0.007	&	0.094	$\pm$	0.003	&	0.087	$\pm$	0.001	\\
HD\,45410 	&	06 30 47.1	&	+58 09 45.4	&	14/01/2011	&	50000	&	0.474	$\pm$	0.034	&	0.198	$\pm$	0.019	&	0.127	$\pm$	0.008	&	0.124	$\pm$	0.007	\\
HD\,68988 	&	08 18 22.1	&	+61 27 38.5	&	23/04/2008	&	50000	&	0.177	$\pm$	0.015	&	0.120	$\pm$	0.007	&	0.092	$\pm$	0.003	&	0.084	$\pm$	0.002	\\
HD\,73534 	&	08 39 15.8	&	+12 57 37.3	&	14/01/2011	&	50000	&	0.287	$\pm$	0.041	&	0.107	$\pm$	0.007	&	0.093	$\pm$	0.004	&	0.093	$\pm$	0.003	\\
HD\,74156 	&	08 42 25.1	&	+04 34 41.1	&	16/01/2009	&	16000	&	0.386	$\pm$	0.040	&	0.216	$\pm$	0.023	&	0.141	$\pm$	0.011	&	0.131	$\pm$	0.009	\\
HD\,75898 	&	08 53 50.8	&	+33 03 24.5	&	23/04/2008	&	50000	&	0.222	$\pm$	0.027	&	0.130	$\pm$	0.010	&	0.096	$\pm$	0.003	&	0.091	$\pm$	0.003	\\
HD\,81688 	&	09 28 39.9	&	+45 36 05.3	&	14/01/2011	&	48149	&	0.647	$\pm$	0.032	&	0.291	$\pm$	0.034	&	0.216	$\pm$	0.024	&	0.216	$\pm$	0.024	\\
HD\,87883 	&	10 08 43.1	&	+34 14 32.1	&	14/01/2011	&	50000	&	0.107	$\pm$	0.004	&	0.084	$\pm$	0.001	&	0.078	$\pm$	0.001	&	0.078	$\pm$	0.001	\\
HD\,90043 	&	10 23 28.3	&	-00 54 08.0	&	14/01/2011	&	50000	&	0.553	$\pm$	0.032	&	0.183	$\pm$	0.017	&	0.125	$\pm$	0.008	&	0.126	$\pm$	0.008	\\
HIP\,57050 	&	11 41 44.6	&	+42 45 07.1	&	14/01/2011	&	50000	&	$<$ 0.072	&		$<$ 0.072		&		$<$ 0.072		&		$<$ 0.072		\\
GJ\,436	&	11 42 11.1	&	+26 42 23.7	&	11/07/2008	&	80000	&	0.080	$\pm$	0.001	&	0.076	$\pm$	0.001	&	0.073	$\pm$	0.001	&		$<$ 0.072		\\
HD\,104985 	&	12 05 15.1	&	+76 54 20.6	&	16/01/2009	&	26480	&	0.682	$\pm$	0.030	&	0.518	$\pm$	0.034	&	0.253	$\pm$	0.025	&	0.238	$\pm$	0.023	\\
HD\,107148 	&	12 19 13.5	&	-03 19 11.2	&	16/01/2009	&	50000	&	0.214	$\pm$	0.021	&	0.147	$\pm$	0.011	&	0.102	$\pm$	0.004	&	0.090	$\pm$	0.002	\\
HD\,109246 	&	12 32 07.1	&	+74 29 22.3	&	14/01/2011	&	50000	&	0.157	$\pm$	0.012	&	0.098	$\pm$	0.003	&	0.087	$\pm$	0.001	&	0.083	$\pm$	0.001	\\
HD\,110014 	&	12 39 14.7	&	-07 59 44.0	&	14/01/2011	&	50000	&	0.881	$\pm$	0.041	&	0.534	$\pm$	0.037	&	0.298	$\pm$	0.037	&	0.287	$\pm$	0.035	\\
HD\,114783 	&	13 12 43.7	&	-02 15 54.1	&	23/04/2008	&	50000	&	0.125	$\pm$	0.006	&	0.086	$\pm$	0.001	&	0.078	$\pm$	0.001	&	0.077	$\pm$	0.001	\\
HD\,115617 	&	13 18 24.3	&	-18 18 40.3	&	14/01/2011	&	50000	&	0.611	$\pm$	0.025	&	0.354	$\pm$	0.031	&	0.158	$\pm$	0.011	&	0.155	$\pm$	0.011	\\
HD\,118203 	&	13 34 02.5	&	+53 43 42.7	&	23/04/2008	&	50000	&	0.249	$\pm$	0.028	&	0.123	$\pm$	0.008	&	0.096	$\pm$	0.003	&	0.092	$\pm$	0.003	\\
HD\,125612\,A 	&	14 20 53.5	&	-17 28 53.4	&	23/04/2008	&	50000	&	0.194	$\pm$	0.021	&	0.130	$\pm$	0.009	&	0.098	$\pm$	0.003	&	0.087	$\pm$	0.002	\\
HD\,128311 	&	14 36 00.6	&	+09 44 47.4	&	14/01/2011	&	50000	&	0.106	$\pm$	0.004	&	0.083	$\pm$	0.001	&	0.077	$\pm$	0.001	&	0.077	$\pm$	0.001	\\
HD\,132406 	&	14 56 54.7	&	+53 22 55.8	&	16/01/2009	&	32000	&	0.200	$\pm$	0.021	&	0.139	$\pm$	0.010	&	0.106	$\pm$	0.005	&	0.093	$\pm$	0.003	\\
HD\,136418 	&	15 19 06.1	&	+41 43 59.5	&	27/07/2011	&	50000	&	0.463	$\pm$	0.041	&	0.178	$\pm$	0.017	&	0.109	$\pm$	0.006	&	0.099	$\pm$	0.004	\\
HD\,137759	&	15 24 55.7	&	+58 57 57.8	&	11/07/2008	&	60000	&	0.709	$\pm$	0.026	&	0.467	$\pm$	0.030	&	0.202	$\pm$	0.017	&	0.191	$\pm$	0.017	\\
HD\,142091 	&	15 51 13.9	&	+35 39 26.6	&	16/01/2009	&	16000	&	0.549	$\pm$	0.027	&	0.422	$\pm$	0.033	&	0.233	$\pm$	0.019	&	0.157	$\pm$	0.011	\\
HD\,145675 	&	16 10 24.3	&	+43 49 03.5	&	11/07/2008	&	80000	&	0.130	$\pm$	0.007	&	0.098	$\pm$	0.002	&	0.083	$\pm$	0.001	&	0.081	$\pm$	0.001	\\
HD\,148427 	&	16 28 28.1	&	-13 23 58.6	&	14/07/2010	&	55000	&	0.371	$\pm$	0.038	&	0.163	$\pm$	0.013	&	0.098	$\pm$	0.003	&	0.093	$\pm$	0.003	\\
HD\,149026 	&	16 30 29.6	&	+38 20 50.3	&	07/09/2009	&	50000	&	0.242	$\pm$	0.025	&	0.128	$\pm$	0.008	&	0.097	$\pm$	0.003	&	0.092	$\pm$	0.003	\\
HD\,150706 	&	16 31 17.5	&	+79 47 23.2	&	07/09/2009	&	50000	&	0.160	$\pm$	0.011	&	0.103	$\pm$	0.004	&	0.086	$\pm$	0.001	&	0.083	$\pm$	0.001	\\
HD\,149143 	&	16 32 51.0	&	+02 05 05.3	&	23/04/2008	&	50000	&	0.225	$\pm$	0.023	&	0.103	$\pm$	0.005	&	0.088	$\pm$	0.002	&	0.087	$\pm$	0.002	\\
GL\,649 	&	16 58 08.8	&	+25 44 38.9	&	14/07/2010	&	40950	&	0.080	$\pm$	0.001	&	0.074	$\pm$	0.001	&		$<$ 0.072		&		$<$ 0.072		\\
HD\,154345 	&	17 02 36.4	&	+47 04 54.7	&	07/09/2009	&	50000	&	0.139	$\pm$	0.009	&	0.098	$\pm$	0.003	&	0.083	$\pm$	0.001	&	0.079	$\pm$	0.001	\\
HD\,155358 	&	17 09 34.6	&	+33 21 21.0	&	14/07/2010	&	53240	&	0.174	$\pm$	0.013	&	0.101	$\pm$	0.003	&	0.086	$\pm$	0.001	&	0.085	$\pm$	0.001	\\
GJ\,1214 	&	17 15 18.9	&	+04 57 49.7 	&	14/07/2010	&	55000	&	0.075	$\pm$	0.001	&		$<$ 0.072		&		$<$ 0.072		&		$<$ 0.072		\\
HD\,156668 	&	17 17 40.4	&	+29 13 38.0	&	14/07/2010	&	55000	&	0.101	$\pm$	0.003	&	0.086	$\pm$	0.001	&	0.078	$\pm$	0.001	&	0.077	$\pm$	0.001	\\
HD\,164922 	&	18 02 30.8	&	+26 18 46.8	&	07/09/2009	&	50000	&	0.133	$\pm$	0.008	&	0.096	$\pm$	0.003	&	0.082	$\pm$	0.001	&	0.080	$\pm$	0.001	\\
HD\,167042 	&	18 10 31.6	&	+54 17 11.5	&	07/09/2009	&	50000	&	0.484	$\pm$	0.028	&	0.256	$\pm$	0.022	&	0.130	$\pm$	0.008	&	0.108	$\pm$	0.005	\\
HD\,170693 	&	18 25 59.1	&	+65 33 48.5	&	14/07/2010	&	55000	&	0.764	$\pm$	0.028	&	0.542	$\pm$	0.029	&	0.279	$\pm$	0.028	&	0.286	$\pm$	0.031	\\
HD\,173416 	&	18 43 36.1	&	+36 33 23.7	&	27/07/2011	&	50000	&	0.741	$\pm$	0.033	&	0.419	$\pm$	0.040	&	0.247	$\pm$	0.027	&	0.242	$\pm$	0.026	\\
HD\,188310 	&	19 54 14.8	&	+08 27 41.2	&	07/09/2009	&	50000	&	0.667	$\pm$	0.030	&	0.455	$\pm$	0.039	&	0.187	$\pm$	0.019	&	0.161	$\pm$	0.014	\\
HD\,200964 	&	21 06 39.8	&	+03 48 11.2	&	27/07/2011	&	50000	&	0.356	$\pm$	0.040	&	0.144	$\pm$	0.012	&	0.116	$\pm$	0.007	&	0.112	$\pm$	0.006	\\
HD\,218566 	&	23 09 10.7	&	-02 15 38.6	&	27/07/2011	&	50000	&	0.107	$\pm$	0.004	&	0.084	$\pm$	0.001	&	0.079	$\pm$	0.001	&	0.078	$\pm$	0.001	\\
HD\,221345 	&	23 31 17.4	&	+39 14 10.3	&	07/09/2009	&	50000	&	0.653	$\pm$	0.028	&	0.365	$\pm$	0.036	&	0.183	$\pm$	0.017	&	0.169	$\pm$	0.014	\\

	\hline\end{tabular}
	\begin{tablenotes}\footnotesize 
	\item[1] the companion candidate (\citealt{b23}) at $0.237\,$arcsec separation could not be resolved
	\end{tablenotes}
	\end{threeparttable}
\end{table*}

\section{Conclusions}
\label{conclusions}
\begin{figure}
\includegraphics[scale=0.4]{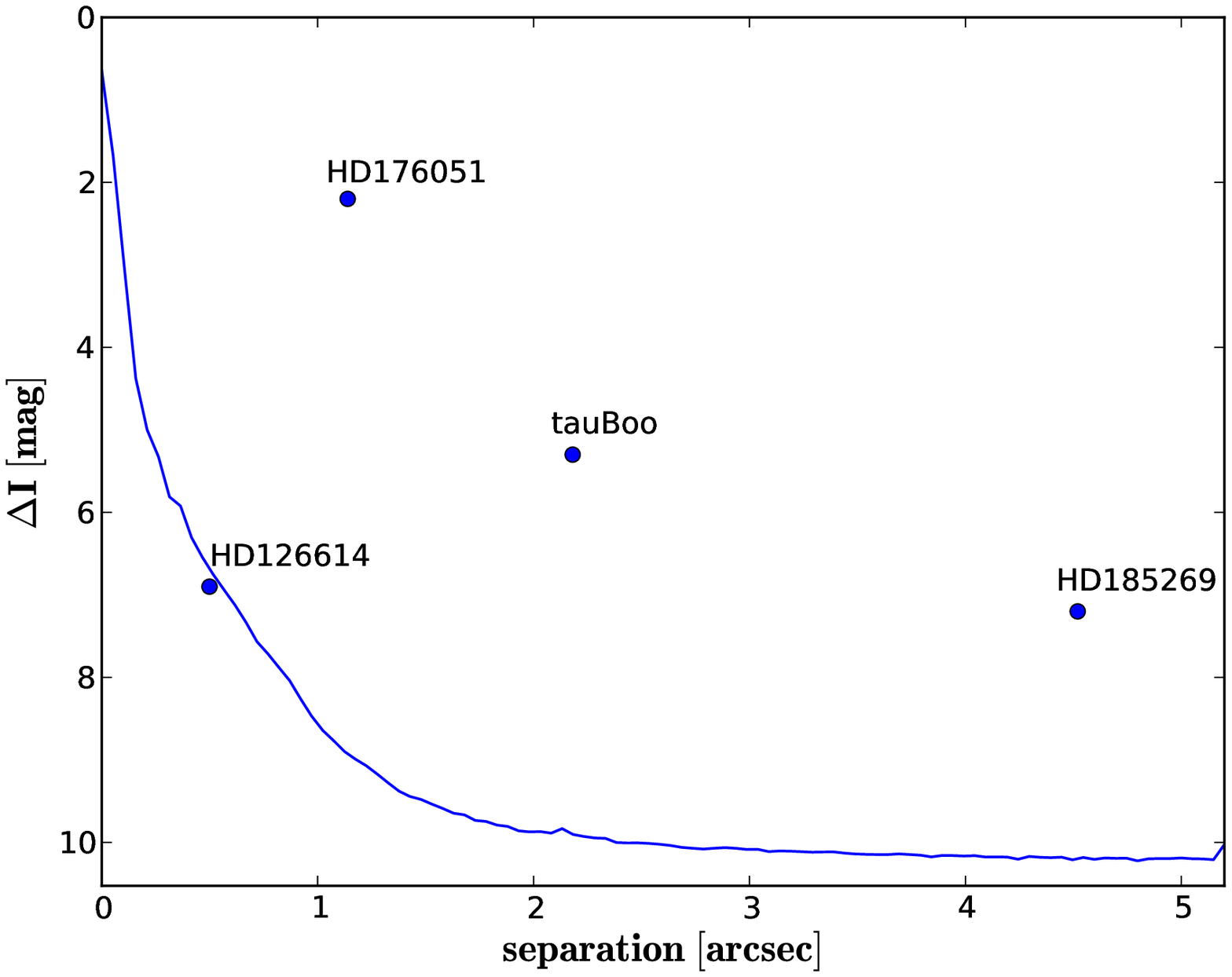}
\caption[]{Average dynamic range in all observation epochs after PSF subtraction. All detected stellar companions have been added. The typical standard deviation is 0.2\,mag.}
\label{fig:average-result}
\end{figure}

In our ongoing study we observed $71$ planet host stars to date. Of these $71$ systems, $3$ were already known to be multiple, for which we present follow up astrometry. Thereby, we show for the first time conclusively, that the companion to HD\,126614 is indeed physically associated with the primary. As the planet host star HD\,126614 also exhibits a further companion at a wider separation (HD\,216614\,C, sep$\,=\,41.914\pm0.110\,$arcsec (3043 AU), PA$\,=\,299.36^\circ\pm0.14^\circ$ at 2MASS Epoch May 3th 2000), the HD\,126614 system is actually a hierarchical triple system, the only one presently known in which the planet host star exhibits a close stellar companion. All other known planet host triples are composed of the planet host star and a binary-companion at wider separation.\\
We also discovered one new low-mass ($0.239 \, \pm \, 0.022 \, M_{\odot}$) stellar companion to the star HD\,185269, with a separation of $4.511\pm 0.013 \,$arcsec at a position angle of $8.44^\circ	\pm 0.30^\circ$. This corresponds to a projected distance of $227\,$AU. HD\,185269A harbors a "Hot Jupiter" ($M\,sini \, = \, 0.94 \, M_{J}$) with an orbital period of $6.8\,$d and a semi-major axis of $0.077 \,$AU, detected by \cite{b20}. They state that its orbital eccentricity of $0.3$ is large in comparison with other planets found within 1\,AU of their host stars. This could possibly be explained by the Kozai mechanism as described by \cite{b31}. We used the formula
$$P \simeq 2 \pi  \sqrt{\frac{a^3_1}{G(m_0 + m_1)}}\left(\frac{m_0 + m_1}{m_2}\right)\left(\frac{a_2}{a_1}\right)^3\left(1-e^2_2\right)^{3/2}$$
to calculate the period of the Kozai oscillations (\citealt{b32}), where the indices 0, 1 and 2 represent the host star, planetary and stellar companions respectively. Assuming an eccentricity of 0.5 for the stellar companion (as suggested by statistical analysis in \citealt{b33}), and a semi-major axis equal to the projected separation of 227\,AU, we get a period of 1.7\,Gyr. Given the age of HD\,185269\,A of 4.2\,Gyr, this period is short enough so that the Kozai effect might have altered the planetary companion's eccentricity and inclination.\\
For the remaining $66$ target stars in our sample we can on average exclude all low-mass stellar companions of $M \,> 0.15\,M_{\odot}$ down to $2\,$arcsec around the primary. For many stars in our sample we get significantly deeper, enabling us to exclude all low-mass stellar companions outside of $2\,$arcsec and up to $12\,$arcsec. To further summarize our detection limits, in Fig.~\ref{fig:average-result} we show the average dynamic range plot of all observations, with all detected stellar companions marked.\\
68 stars of our sample had an unknown multiplicity status before our observations. Of those, only one proved to be multiple within the detection limits of our survey. This yields a multiplicity rate of only 1.5\,\% for our sample. Given that the multiplicity rate of the known exoplanet host population is about 17\,\%, stellar multiple systems are underrepresented in our sample. However, since the sample size of our Astralux survey is not yet statistically significant ($\geq$ 231 for 95\,\% confidence level and an error margin of 5\,\%) we can not draw any conclusion for the whole exoplanet population. For a detailed statistical analyzis of the properties of all known stellar multiple systems harboring extrasolar planets, we refer to \cite{b30}.\\ 
We will continue our current monitoring campaign in order to determine the multiplicity status of all exoplanet host stars, which then will yield the true multiplicity rate of planet-bearing stars. This will also eventually allow us to draw conclusions about the frequency of planets in multiple stellar systems, as well as their properties compared to planets which reside around single stars. 

\section*{Acknowledgments}
CG and TE wish to acknowledge Deutsche Forschungsgemeintschaft (DFG) for grant NE 515 / 30-1.
MS would like to thank DFG for support in project NE 515 / 36-1.
Based on observations collected at the Centro Astron\'{o}mico Hispano Alem\'{a}n (CAHA) at Calar Alto, 
operated jointly by the Max-Planck Institut f\"{u}r Astronomie and the Instituto de Astrof\'{i}sica de Andalucía (CSIC). 
We would especially like to express our thanks to the very helpful staff at the Calar Alto observation site.
We used Simbad and Vizier as well as archival data from HST.
HST data were obtained from the data archive at the Space Telescope Institute, 
which is operated by the association of Universities for Research in Astronomy, Inc. under the NASA contract NAS 5-26555.
This publication makes use of data products from the Two Micron All Sky Survey, which is a joint project of the University of Massachusetts and the Infrared Processing and Analysis Center/California Institute of Technology, funded by the National Aeronautics and Space Administration and the National Science Foundation.
This research has made use of the Washington Double Star Catalog maintained at the U.S. Naval Observatory.
We made use of the SciPy tools for scientific computation in Python by \cite{b21}. 
CG would also like to express special thanks to Tristan R\"oll and Christian Adam for their invaluable help and comments regarding the Python programming language as well as Tobias Schmidt and Hiroshi Kobayashi for fruitful discussion.
Finally, we would like to thank Donna Keeley for the language editing of our manuscript.

\label{lastpage}

\end{document}